\documentclass[journal]{IEEEtran}

\usepackage{amsfonts}
\usepackage{amssymb}
\usepackage{cite}
\usepackage[cmex10]{amsmath}
\ifCLASSINFOpdf
\usepackage[pdftex]{graphicx}
\graphicspath{{./Images/}}
\usepackage{algorithmic}
\usepackage{algorithm}
\usepackage{array}
\usepackage{pgfplots}
\usepackage{pgfplotstable}
\newcounter{marknumber}

\pgfplotsset{
    error bars/every nth mark/.style={
        /pgfplots/error bars/draw error bar/.prefix code={
            \pgfmathtruncatemacro\marknumbercheck{mod(floor(\themarknumber/2),#1)}
            \ifnum\marknumbercheck=0
            \else
                \begin{scope}[opacity=0]
            \fi
        },
        /pgfplots/error bars/draw error bar/.append code={
            \ifnum\marknumbercheck=0
            \else
                \end{scope}
            \fi
            \stepcounter{marknumber}    
        }
    }
}
\usepgfplotslibrary{external} 
\pgfplotsset{compat=newest}

\DeclareMathOperator*{\argmax}{arg\,max}
\ifCLASSOPTIONcompsoc
    \usepackage[caption=false, font=normalsize, labelfont=sf, textfont=sf]{subfig}
\else
\usepackage[caption=false, font=footnotesize]{subfig}
\fi

\IEEEoverridecommandlockouts
\usepackage{xcolor}
\usepackage{amssymb}
\def\BibTeX{{\rm B\kern-.05em{\sc i\kern-.025em b}\kern-.08em
    T\kern-.1667em\lower.7ex\hbox{E}\kern-.125emX}}

\begin{document}
\title{\textit{DeepWiVe}: Deep-Learning-Aided \\ Wireless Video Transmission}
%
\author{Tze-Yang Tung and Deniz G\"und\"uz \\
Information Processing and Communications Lab (IPC-Lab),\\ Imperial College London, UK \\
\texttt{\{tze-yang.tung14, d.gunduz\}@imperial.ac.uk}
\thanks{This work was supported by the European Research Council (ERC) through project BEACON (No. 677854).}
}

\maketitle

\begin{abstract}
We present \textit{DeepWiVe}, the first-ever end-to-end joint source-channel coding (JSCC) video transmission scheme that leverages the power of deep neural networks (DNNs) to directly map video signals to channel symbols, combining video compression, channel coding, and modulation steps into a single neural transform. 
Our DNN decoder predicts residuals without distortion feedback, which improves video quality by accounting for occlusion/disocclusion and camera movements. 
We simultaneously train different bandwidth allocation networks for the frames to allow variable bandwidth transmission. 
Then, we train a bandwidth allocation network using reinforcement learning (RL) that optimizes the allocation of limited available channel bandwidth among video frames to maximize overall visual quality. 
Our results show that \textit{DeepWiVe} can overcome the \textit{cliff-effect}, which is prevalent in conventional separation-based digital communication schemes, and achieve graceful degradation with the mismatch between the estimated and actual channel qualities.
\textit{DeepWiVe} outperforms H.264 video compression followed by low-density parity check (LDPC) codes in all channel conditions by up to 0.0462 on average in terms of the multi-scale structural similarity index measure (MS-SSIM), while beating H.265 + LDPC by  up to 0.0058 on average.
We also illustrate the importance of optimizing bandwidth allocation in JSCC video transmission by showing that our optimal bandwidth allocation policy is superior to the na\"ive uniform allocation. 
We believe this is an important step towards fulfilling the potential of an end-to-end optimized JSCC wireless video transmission system that is superior to the current separation-based designs.
\end{abstract}

\section{Introduction}
\label{sec:intro}

\begin{figure}
    \centering
    \includegraphics[width=\linewidth]{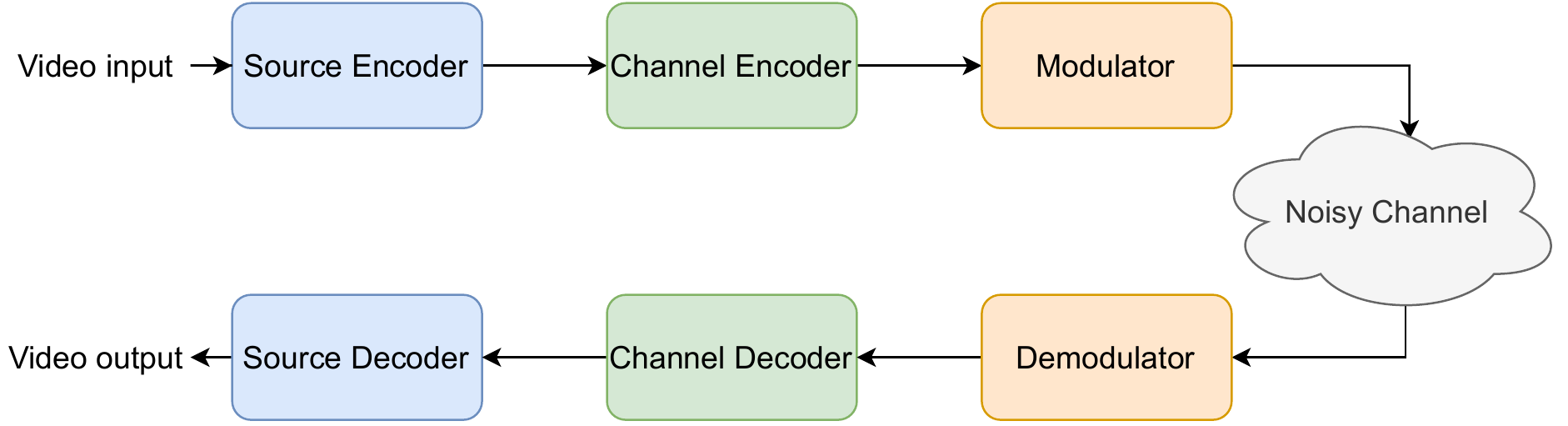}
    \caption{Diagram of a typical separation-based digital video delivery system employed by almost all communication systems today.}
    \label{fig:separation}
\end{figure}

Video content contributes to more than 80\% of Internet traffic and the percentage is only expected to increase \cite{noauthor_cisco_2017}.
Video compression is widely used to reduce the bandwidth requirement when transmitting video signals wirelessly.
This follows the modular approach employed in almost all wireless video transmission systems, where the end-to-end transmission problem is divided into two: (1) a source encoder that compresses the video into a sequence of bits of the shortest possible length such that a reconstruction of the original video is possible within an allowable distortion; and 
(2) a channel encoder that introduces redundancies such that the compressed bits are protected against channel errors and interference.
A diagram of this modular-based approach is shown in Fig. \ref{fig:separation}.

This separate source and channel coding design provides modularity and allows independent optimization of each component. 
It has been applied successfully in a large variety of applications from on-demand mobile video streaming to video conferencing and digital TV broadcasting.
However, the limits of the separation-based designs are beginning to rear, with the emergence of more demanding and challenging video delivery applications, such as wireless virtual reality (VR) and drone-based surveillance systems, which have ultra-low latency requirements, suffer from highly unpredictable channel conditions, and need to be implemented on energy limited mobile devices.

In the context of wireless video transmission, separation-based designs lead to what is known as the \textit{cliff-effect}. 
That is, when the channel condition deteriorates below the level anticipated by the channel encoder, the source information becomes irrecoverable. 
This leads to a cliff edge deterioration of the system performance. 
As a result, most current systems operate at a much more conservative transmission rate than that is suggested by the instantaneous channel capacity, and employ additional error correction mechanisms through automatic repeat requests (ARQ). 

An alternative to the separation-based architecture is joint source-channel coding (JSCC).
The most natural JSCC approach continues to employ separate modules for compression and communication, but jointly optimizes various parameters of these modules in a cross-layer framework.
While there have been many such proposals over the years \cite{cheung_bit_2000,kozintsev_robust_1998,kondi_joint_2001,ji_joint_2012,cheung_joint_1997,kondi_joint_2002,chen_adaptive_2002,wu_joint_2013}, these techniques typically do not provide sufficient gains to justify the significant increase in system complexity.

A more fundamental approach is to design the transmission system from scratch, without considering any digital interface in between.
The best example for such an approach is analog communications, such as AM/FM radio or analog TV, where the information is directly modulated onto the carrier waveform without any compression. 
Analog communication can overcome the cliff-effect problem by showing graceful degradation with the channel parameters.
From a fundamental information theoretic perspective, when transmitting  independent Gaussian samples over an additive white Gaussian noise (AWGN) channel, with one sample per channel use on average, uncoded transmission, where each sample is simply scaled and transmitted, meets the theoretical Shannon bound \cite{goblick_coding_1969}.
With digital transmission, the same performance can only be achieved by vector-quantizing an arbitrarily long sequence of source samples, followed by a capacity achieving channel code.
Benefits of analog transmission has also been shown in various multi-user scenarios \cite{aguerri_joint_2016, lapidoth_sending_2007}. 
However, analog modulation cannot exploit the available bandwidth efficiently, and the optimality of simple uncoded transmission does not generalize to bandwidth-mismatched scenarios.
However, despite this theoretical suboptimality, analog modulation approaches to image and video transmission have gained recent popularity \cite{jakubczak_softcast:_2010, tung_sparsecast:_2018, yin_compressive_2016,wang_wireless_2014}, mainly due to their low computational complexity and the graceful degradation with channel quality.

Recently, it has been shown in \cite{bourtsoulatze_deep_2019, Bourtsoulatze:TCCN:19} that deep neural networks (DNNs) can break the complexity barrier in designing effective JSCC schemes, focusing particularly on the image transmission problem.
The approach there is to train DNN models in an autoencoder architecture with a non-trainable channel layer between the encoder and decoder. 
The authors showed that this approach not only provides graceful degradation with channel quality, but can also achieve results similar to or superior than state-of-the-art separation-based digital designs.
An extension to this work \cite{kurka_bandwidth-agile_2021} further shows that JSCC is successively refineable; that is, an image can be transmitted in stages, and each additional channel block further refines the quality of the decoded image, with almost no additional cost.


Herein, we propose a deep learning-based JSCC solution for wireless video transmission, called \textit{DeepWiVe}, which is trained to optimize the reconstructed video quality in an end-to-end fashion. 
Although the problem of video compression using deep learning has received significant attention \cite{wu_video_2018,lu_dvc:_2018,rippel_learned_2018,djelouah_neural_2019,habibian_video_2019,han_deep_2019,agustsson_scale-space_2020,liu_deep_2021,hu_fvc_2021}, no prior work has considered deep learning aided wireless video delivery.

\begin{figure}
    \centering
    \includegraphics[width=\linewidth]{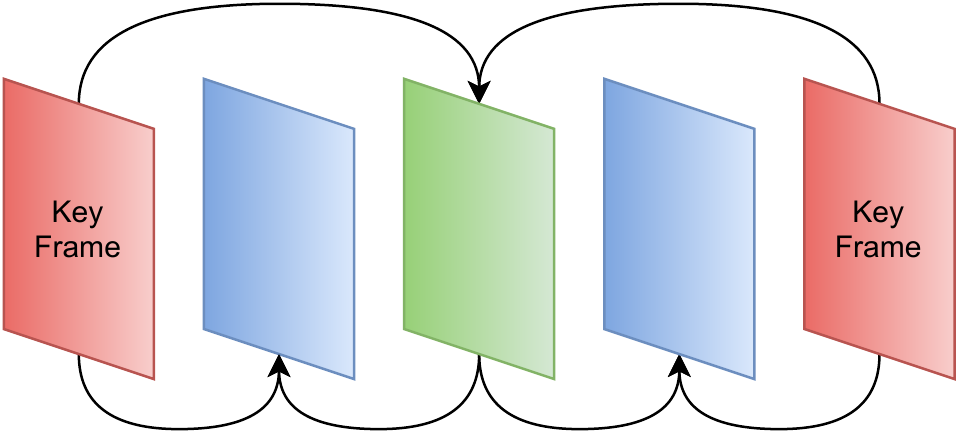}
    \caption{Diagram of a typical interpolation structure used in video compression algorithm.}
    \label{fig:codec_interp_structure}
\end{figure}

The core idea behind most video compression algorithms is to exploit the temporal correlations across video frames.
In a standard video sequence, motion differences between successive frames is typically very small, and video compression algorithms can be very efficient by identifying intermittent key frames, which are compressed and decoded independently, and conveying only the motion and residual information for the remaining frames, thereby exploiting temporal redundancy.
Here, the residual information refers to the difference between the true frame and the motion compensated key frame. 
Fig. \ref{fig:codec_interp_structure} shows a typical interpolation structure in a video compression algorithm.
For compression algorithms, the residual information is available at the encoder as the encoder can simply decode the compressed frames and observe the difference between the source and its reconstruction.
On the other hand, for JSCC, the residual depends on the channel condition during transmission.
Therefore, the residual is not known at the encoder. 
To overcome this, we propose to use a DNN to predict the residual, without the need for distortion feedback.

In \textit{DeepWiVe} we directly map the video sequence into the channel vector under a channel bandwidth constraint for the transmission of a group-of-pictures (GoP). 
Similarly to \cite{kurka_bandwidth-agile_2021}, we employ variable bandwidth transmission and allocate the bandwidth dynamically using reinforcement learning (RL) to train a bandwidth allocation network that optimizes the bandwidth utilization of each frame. 
Our results show that \textit{DeepWiVe} can meet or beat industry standard video compression codecs, such as H.264, combined with state-of-the-art channel codes, such as low density parity check (LDPC) codes, in almost all channel conditions tested, while achieving graceful degradation of video quality with respect to channel quality, thereby avoiding the \textit{cliff-effect}.

\begin{figure}
    \centering
    \includegraphics[width=\linewidth]{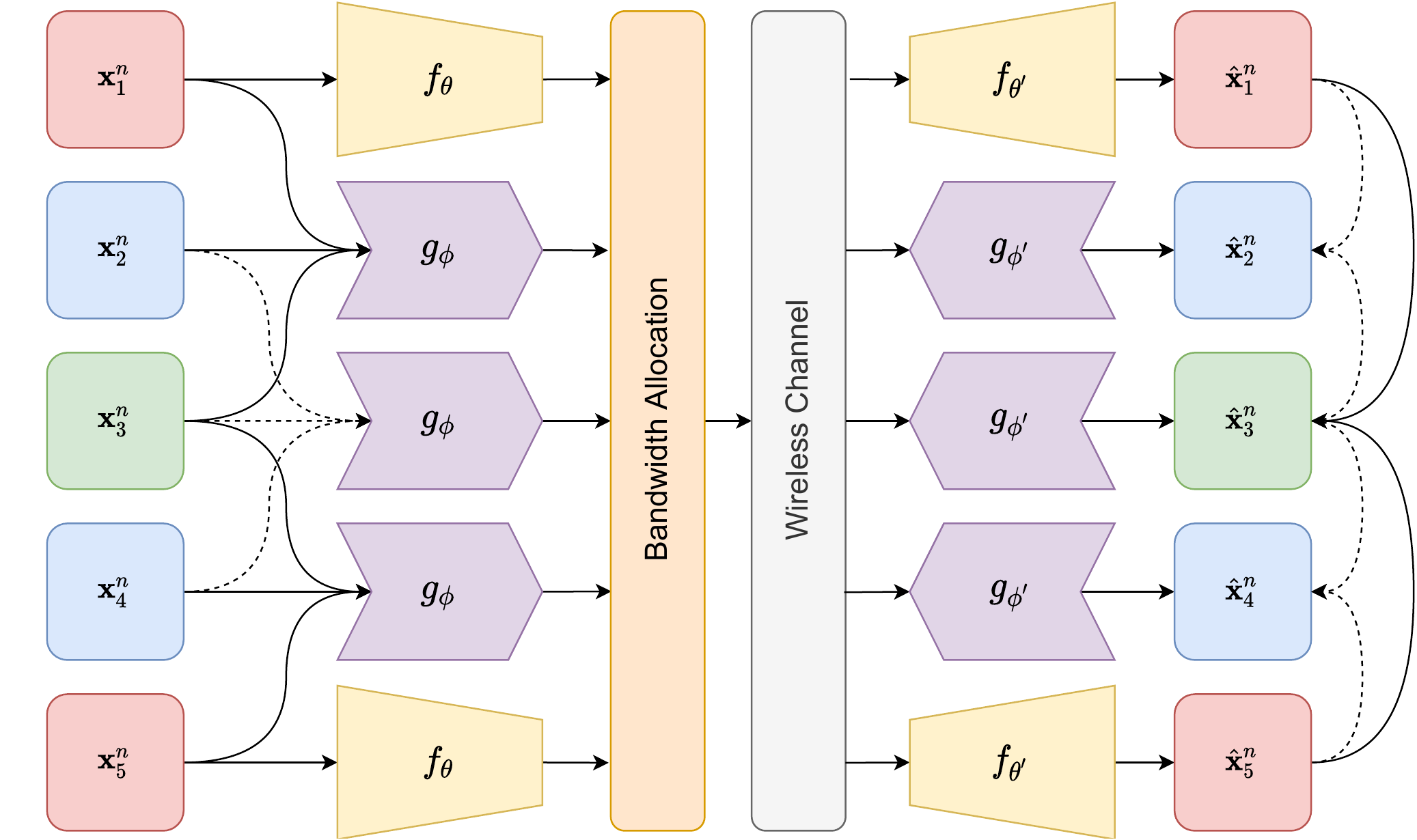}
    \caption{\textit{DeepWiVe} system overview.}
    \label{fig:sys_overview}
    \vspace{-0.4cm}
\end{figure}

The contributions of this paper are summarized as follows: 
\begin{enumerate}
    \item We propose \textit{DeepWiVe}, a JSCC-based wireless video transmission scheme leveraging DNNs to jointly compress and channel code video frames in an end-to-end manner to maximize the end video quality.
    \item We train our DNN decoder to predict residuals without the need for distortion feedback.
    \item We optimize the bandwidth allocation among video frames by training a bandwidth allocation network using RL, and show that doing so achieves superior performance than na\"ive uniform allocation.
    \item We show that it is possible to achieve variable bandwidth transmission by simply training different bandwidth allocation networks.
    \item Numerical results show that our proposed \textit{DeepWiVe} is superior to industry standard H.264 \cite{wiegand_overview_2003} codec with state-of-the-art LDPC channel codes \cite{gallager_low-density_1962} in all channel conditions tested and can avoid the cliff-effect.
    It also beats H.265 \cite{ohm_high_2013} using the same channel codes when evaluated in terms of the multi-scale structural similarity index measure (MS-SSIM).
\end{enumerate}

\section{Related Work}
\label{sec:related_work}

JSCC for video delivery has consistently received attention over the years.
The earliest work we could find is \cite{cheung_joint_1997}, which studies the problem of video multicast to heterogeneous receivers. 
They approached the problem from the receivers' perspective, where the source video is encoded in a hierarchical manner, with each layer of the hierarchy distributed on a separate network channel. 
Each receiver then adapts to its local channel capacity by adjusting the number of layers it decodes.
In a similar line of work, \cite{stoufs_scalable_2008} uses scalable video coding (SVC), which encodes the source video into multiple bitstreams, with a base layer that represents the lowest supported quality and a set of enhancement layers representing versions of the video at different qualities. 
Since the base layer and lower quality layers require more error protection than higher quality layers, unequal error protection (UEP) of packets is used.
To determine the optimal channel code rate for each layer such that the average distortion is minimized, they devised a low complexity search algorithm to find the optimal choice of channel code rates among a set of available rates. 
There is a large body of works that use source coding schemes similarly to SVC and minimize the end-to-end distortion by jointly optimizing various parameters of the source and channel codes \cite{kondi_joint_2001, ji_joint_2012, kondi_joint_2002, chen_adaptive_2002,wu_joint_2013, Shutoy:JSTSP:07}.
However, none of the scheme proposed were able to achieve adequate performance gains for the increased complexity they introduce as a result of the optimization of the parameters.
Moreover, with the exception of \cite{cheung_joint_1997} and \cite{chen_adaptive_2002}, which model the channel as a physical link with a certain error probability, the above works mainly consider time-varying channels with proposed schemes aimed at adapting to channel variations.
In this work, we will focus on AWGN channels.

A completely refreshed approach to JSCC video delivery, called SoftCast, utilizing low complexity methods to map videos or images from the pixel domain to channel symbols directly was first introduced in \cite{jakubczak_softcast:_2010}. 
Their scheme involves a hybrid digital and analog design by leveraging frequency domain sparsity. 
Since then, various works have improved upon \cite{jakubczak_softcast:_2010} by optimizing different aspects of the hybrid digital and analog design.
In \cite{jakubczak_softcast:_2010, tung_sparsecast:_2018, yin_compressive_2016,wang_wireless_2014}, frequency domain sparsity is exploited by utilizing compressed sensing to reduce the bandwidth requirement.
In \cite{xiong_power-distortion_2013}, the power allocation problem is addressed by optimizing the division of frequency domain coefficients into chunks.
In \cite{song_hybridcast_2014}, the proposed scheme separates a base layer (low frequency coefficients) from the enhancement layer (high frequency coefficients) and sends the base layer using a digital separation-based system, while the enhancement layer is sent using a scheme similar to SoftCast.
This method also improves power allocation since low frequency coefficients have larger values than high frequency coefficients in natural images so the majority of the signal energy is sent through the digital system.
In \cite{xiong_g-cast_2014, liu_cg-cast_2019}, a similar approach is utilized, however, instead of the high frequency components, they send the gradient of the image as the enhancement information.
In \cite{xiong_high_2014}, the prior used for the estimation is modified to obtain better reconstruction quality (SoftCast assumes a Gaussian prior).
In \cite{trioux_performance_2021}, the variability of temporal redundancies between frames is addressed by introducing adaptive GoP sizes.
Although these methods have been shown to overcome the \textit{cliff-effect}, they are not competitive to separation-based schemes when it comes to video quality and cannot exploit the available bandwidth, or adapt to channel and network conditions dynamically. 

Following the success of DNNs in various image and video intelligence tasks, there has been a growing interest in employing them for video compression \cite{agustsson_scale-space_2020,wu_video_2018,lu_dvc:_2018,rippel_learned_2018,djelouah_neural_2019,habibian_video_2019,han_deep_2019,hu_fvc_2021,liu_deep_2021}. 
Although some recent works have reported competitive or superior performance with respect to state-of-the-art video compression standards H.264/5, none of the works have considered the wireless video delivery problem, taking into account the channel conditions. In principle, the works treat the communication layer simply as a perfect bit pipe.
As a result, the shortcomings of separation-based schemes, such as the \textit{cliff-effect}, are inherent to those works if applied in a wireless communication scenario.

The closest prior art to our work are \cite{bourtsoulatze_deep_2019,Bourtsoulatze:TCCN:19, xu_wireless_2020,kurka_bandwidth-agile_2021,kurka_deepjscc-f_2020}, which explore the JSCC problem, but they focus on image transmission.
In the context of video transmission, there are unique challenges that sets it apart from simple image transmission.
Namely, exploiting the inter-frame redundancies to improve coding efficiency and to optimize resource allocation across the frames.
Therefore, extending the problem from image to video transmission is not a trivial task.

\begin{figure*}
    \centering
    \includegraphics[width=\textwidth]{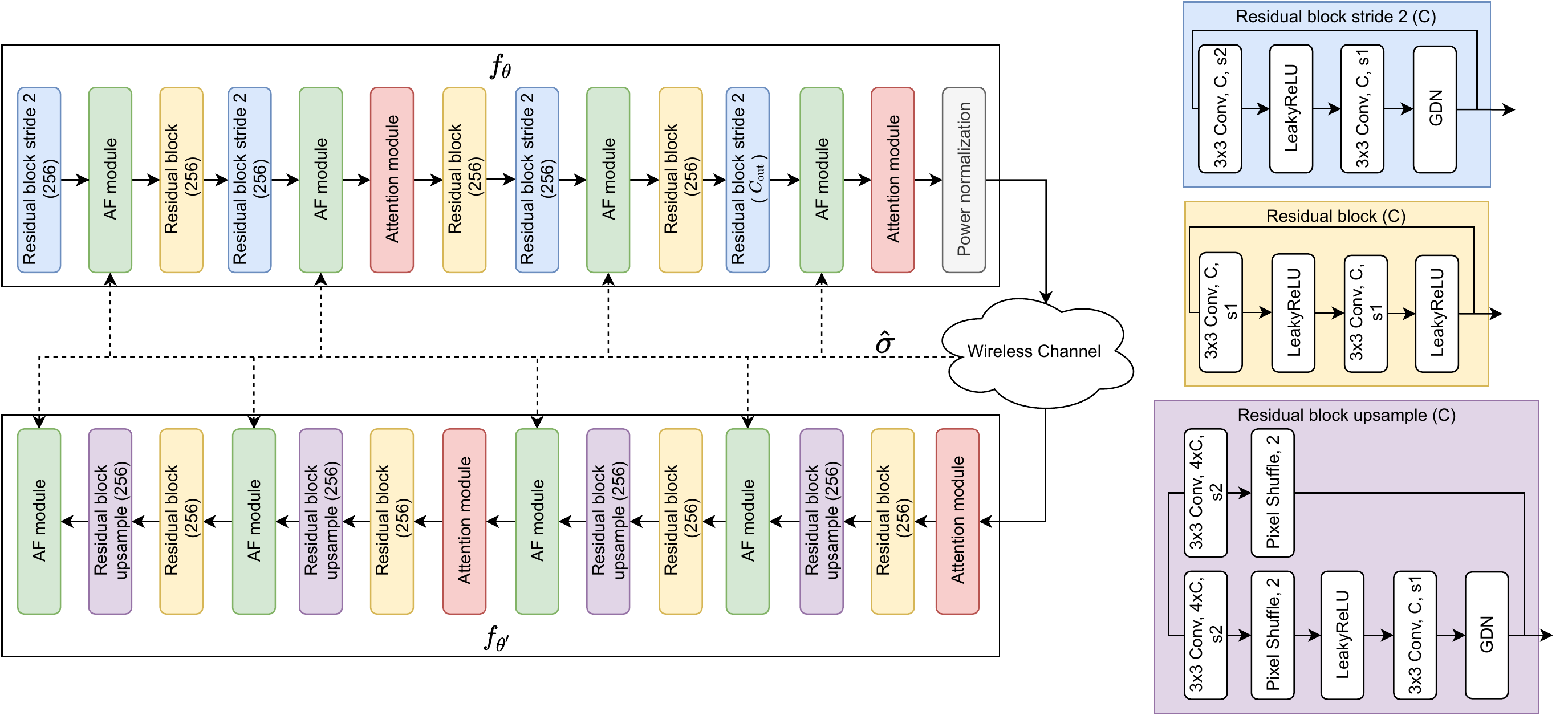}
    \caption{Key frame encoder/decoder $(f_{\boldsymbol{\theta}},f_{\boldsymbol{\theta}^\prime})$ network architectures.}
    \label{fig:anchor_net}
\end{figure*}

\section{Problem Formulation}
\label{sec:problem_form}

We consider the problem of wireless video transmission in a constrained bandwidth setting.
Consider a video sequence $\mathbf{X}=\{\mathbf{X}^n\}_{n=1}^T$, where $\mathbf{X}^n=\{\mathbf{x}_1^n,\dots,\mathbf{x}_N^n\},~\mathbf{x}_i^n\in\mathbb{R}^{H\times W\times 3},~\forall i\in [1,N]$, represents the $n$th GoP in the video sequence. 
Each frame $\mathbf{x}_i^n$ is represented as a 24bit RGB image.
We wish to design an encoding function $E:\mathbb{R}^{TN\times H\times W\times 3}\mapsto \mathbb{C}^{Tk}$, which maps the video sequence $\mathbf{X}$ to a set of complex symbols $\mathbf{z} = E(\mathbf{X})\in\mathbb{C}^{Tk}$, and a decoding function $D:\mathbb{C}^{Tk} \mapsto \mathbb{R}^{TN\times H\times W\times 3}$, which  maps a noise corrupted version of the encoder output $\mathbf{y}=\mathbf{z}+\mathbf{n}$, caused by the wireless channel, to an approximate reconstruction of the original video sequence $\hat{\mathbf{X}}=D(\mathbf{y})$.

In this setting, we restrict the number of channel uses to $k$ per GoP, which can be considered as a bandwidth constraint and we define the \textit{bandwidth compression ratio} as
\begin{equation}
    \rho = \frac{k}{3HWN}.
\end{equation}

We consider an additive white Gaussian noise (AWGN) channel $\mathbf{n}\sim\text{CN}(0,\sigma^2\mathbf{I})$ that follows a complex Gaussian noise distribution with zero mean and covariance $\sigma^2\mathbf{I}$ ($\mathbf{I}$ is the identity matrix).
We impose an average power constraint at the transmitter, such that
\begin{equation}
    \frac{1}{Tk}\mathbb{E}_{\mathbf{z}}\left[||\mathbf{z}||_2^2\right]\leq P,
\end{equation}
where the expectation is taken over the distribution of the encoder output.
Accordingly, the channel signal-to-noise ratio (SNR) is defined as
\begin{equation}
    \text{SNR}=10\log_{10}\bigg(\frac{P}{\sigma^2}\bigg)~\text{dB}.
\end{equation}

We measure the average quality of the reconstructed video using two metrics: peak signal-to-noise ratio (PSNR) and MS-SSIM.
They are defined as 
\begin{equation}
    \text{PSNR}(\mathbf{X},\hat{\mathbf{X}})=\frac{1}{TN}\sum_{n=1}^T\sum_{i=1}^N 10\log_{10}\bigg(\frac{255^2}{l_{\text{PSNR}}(\mathbf{x}_i^n,\hat{\mathbf{x}}_i^n)}\bigg)~\text{dB},
    \label{eq:psnr_avg}
\end{equation}
and
\begin{equation}
    \text{MS-SSIM}(\mathbf{X},\hat{\mathbf{X}})=\frac{1}{TN}\sum_{n=1}^T\sum_{i=1}^N 1 - l_{\text{MS-SSIM}}(\mathbf{x}_i^n,\hat{\mathbf{x}}_i^n),
    \label{eq:msssim_avg}
\end{equation}
where 
\begin{equation}
    l_{\text{PSNR}}(\mathbf{x}_i^n,\hat{\mathbf{x}}_i^n) = \frac{1}{3HW}||\mathbf{x}_i^n - \hat{\mathbf{x}}_i^n||_2^2,
    \label{eq:mse_loss}
\end{equation}
and
\begin{equation}
    l_{\text{MS-SSIM}}(\mathbf{x}_i^n,\hat{\mathbf{x}}_i^n) = 1 - \text{MS-SSIM}(\mathbf{x}_i^n,\hat{\mathbf{x}}_i^n).
    \label{eq:msssim_loss}
\end{equation}
MS-SSIM is defined between two frames as
\begin{align}
    &\text{MS-SSIM}(\mathbf{x}_i^n,\hat{\mathbf{x}}_i^n)\\
    &= [l_M(\mathbf{x}_i^n,\hat{\mathbf{x}}_i^n)]^{\alpha_M}\prod_{j=1}^M[c_j(\mathbf{x}_i^n,\hat{\mathbf{x}}_i^n)]^{\beta_j}[s_j(\mathbf{x}_i^n,\hat{\mathbf{x}}_i^n)]^{\gamma_j},
\end{align}
where
\begin{align}
    l_M(\mathbf{x}_i^n,\hat{\mathbf{x}}_i^n) &= \frac{2\mu_{\mathbf{x}_i^n}\mu_{\hat{\mathbf{x}}_i^n}+c_1}{\mu_{\mathbf{x}_i^n}^2+\mu_{\hat{\mathbf{x}}_i^n}^2+c_1},\\
    c_j(\mathbf{x}_i^n,\hat{\mathbf{x}}_i^n) &= \frac{2\sigma_{\mathbf{x}_i^n}\sigma_{\hat{\mathbf{x}}_i^n}+c_2}{\sigma_{\mathbf{x}_i^n}^2+\sigma_{\hat{\mathbf{x}}_i^n}^2+c_2},\\
    s_j(\mathbf{x}_i^n,\hat{\mathbf{x}}_i^n) &= \frac{\sigma_{\mathbf{x}_i^n\hat{\mathbf{x}}_i^n}+c_3}{\sigma_{\mathbf{x}_i^n}\sigma_{\hat{\mathbf{x}}_i^n}+c_3}.
\end{align}
Here, $\mu_{\mathbf{x}_i^n}$, $\sigma^2_{\mathbf{x}_i^n}$, $\sigma^2_{\mathbf{x}_i^n\hat{\mathbf{x}}_i^n}$ are the mean and variance of $\mathbf{x}_i^n$, and the covariance between $\mathbf{x}_i^n$ and $\hat{\mathbf{x}}_i^n$, respectively.
$c_1$, $c_2$, and $c_3$ are coefficients for numeric stability; 
$\alpha_M$, $\beta_j$, and $\gamma_j$ are the weights for each of the components.
Each $c_j(\cdot)$ and $s_j(\cdot)$ are computed at a different downsampled scale of $(\mathbf{x}_i^n,\hat{\mathbf{x}}_i^n)$.
We use the default parameter values of ($\alpha_M$, $\beta_j$, $\gamma_j$) provided by the original paper \cite{wang_multiscale_2003}.
MS-SSIM has been shown to be as good and better at approximating human visual perception than the more simplistic structural similarity index (SSIM) on different subjective image and video databases.

The overall goal of our design is to maximize the video quality, measured by either Eqn. (\ref{eq:psnr_avg}) or (\ref{eq:msssim_avg}), between the input video $\mathbf{X}$ and its reconstruction at the decoder $\hat{\mathbf{X}}$, under the given constraints on the available bandwidth ratio $\rho$ and the average power $P$.

\subsection{Joint Source-Channel Video Coding}
\label{subsec:video_jscc}

In this section, we present our proposed DNN-based joint source-channel video encoding and decoding scheme.
We will deconstruct the design of the encoder ($E$) and decoder ($D$) into three parts: the key frame encoder/decoder ($f_{\boldsymbol{\theta}},f_{\boldsymbol{\theta}^\prime}$), parameterized by ($\boldsymbol{\theta},\boldsymbol{\theta}^\prime$), the interpolation encoder/decoder ($g_{\boldsymbol{\phi}},g_{\boldsymbol{\phi}^\prime}$), parameterized by ($\boldsymbol{\phi},\boldsymbol{\phi}^\prime$), and the bandwidth allocation function $q_{\boldsymbol{\psi}}$, parameterized by $\boldsymbol{\psi}$.
We will represent all these functions with DNNs, where the parameters of the functions correspond to the weights of these DNNs.
An overview of our scheme is shown in Fig. \ref{fig:sys_overview}.

One of the drawbacks in prior works was the need to train multiple networks, one for each channel condition.
To address this issue, \cite{xu_wireless_2020} proposed an attention feature (AF) module, motivated by resource assignment strategies in traditional JSCC schemes \cite{sayood_joint_2000}, which allows the network to learn to assign different weights to different features for a given SNR.
This is achieved by deliberately randomizing the channel SNR during training, and providing the AF modules with the current SNR.
By doing so, the results in \cite{xu_wireless_2020} show that a single model whose parameters are adjusted to the channel SNR with the help of the AF modules perform at least as well as the models trained for each SNR individually.
We adopt the AF module proposed by \cite{xu_wireless_2020} in \textit{DeepWive} to obtain a single model that can work over a range of SNRs. 

The encoding and decoding procedures are described herein.
Consider the $n$th GoP, $\mathbf{X}^n=\{\mathbf{x}_1^n,...,\mathbf{x}_N^n\}$. 
The last ($\mathbf{x}_N^n$) frame is called the key frame and is compressed and transmitted using the key frame encoder $f_{\boldsymbol{\theta}}:\mathbb{R}^{H\times W\times 3}\mapsto\mathbb{C}^k$,
\begin{equation}
    \mathbf{z}_i^n = f_{\boldsymbol{\theta}}(\mathbf{x}_i^n, \hat{\sigma}^2),~i=N,
    \label{eq:img_encoder}
\end{equation}
where $\hat{\sigma}^2$ is the estimated channel noise power at the transmitter.
Here, each element of $\mathbf{z}_i^n$, denoted by $z_{i,j}^n$, represents the in-phase (I) and quadrature (Q) components of a complex channel symbol.
We note that, while the channel input/output values are complex, we employ real-valued DNN architecture. 
The mapping between real network outputs and complex channel inputs (or vice versa) is achieved by pairing consecutive real values at the output of the encoder DNN.

The values in the complex latent vector are first normalized according to
\begin{equation}
    \hat{z}_{i,j}^n = \sqrt{kP}\frac{z_{i,j}^n}{\sqrt{(\mathbf{z}^n_i)^H\mathbf{z}^n_i}},~j=1,\dots,k,
    \label{eq:power_norm}
\end{equation}
where $P$ is the power constraint, $k$ is the bandwidth constraint, and $H$ refers to the Hermitian transpose. 

These values are then directly sent through the channel as,
\begin{equation}
    \hat{\mathbf{y}}_i^n = \hat{\mathbf{z}}_i^n + \mathbf{n}.
    \label{eq:awgn_channel}
\end{equation}
Consequently, the key frame decoder $f_{\boldsymbol{\theta}^\prime}:\mathbb{C}^k\mapsto\mathbb{R}^{H\times W\times 3}$ that maps the noisy latent vector $\hat{\mathbf{y}}_i^n\in\mathbb{C}^k$ observed at the receiver back to the original frame domain $\hat{\mathbf{x}}_i^n\in\mathbb{R}^{H\times W\times 3}$ is defined as:
\begin{equation}
    \hat{\mathbf{x}}_i^n = f_{\boldsymbol{\theta}^\prime}(\hat{\mathbf{y}}_i^n, \hat{\sigma}^2),~i=N.
    \label{eq:img_decoder}
\end{equation}
The loss between the original frame $\mathbf{x}_i^n$ and the reconstructed frame $\hat{\mathbf{x}}_i^n$ is computed using Eqn. (\ref{eq:mse_loss}) or (\ref{eq:msssim_loss}) depending on which performance measure is being used.
The network weights $(\boldsymbol{\theta},\boldsymbol{\theta}^\prime)$ are then updated via backpropagation with respect to the gradient of the loss.

The network architectures of the key frame encoder and decoder are shown in Fig. \ref{fig:anchor_net}.
The Pixel Shuffle module, first proposed in \cite{shi_real-time_2016}, increases the height and width of the input tensor dimensions efficiently by exchanging channel dimensions for height and width dimensions.
The GDN layer refers to Generalised Divisive Normalization, initially proposed in \cite{balle_density_2016}, and has been shown to be effective in density modeling and compression of images.
The Attention layer refers to the simplified attention module proposed in \cite{cheng_learned_2020}, which reduces the computation cost of the attention module originally proposed in \cite{wang_non-local_2018}.
The attention mechanism has been used in both \cite{cheng_learned_2020} and \cite{wang_non-local_2018} to improve the compression efficiency by focusing the neural network on regions in the image that require higher bit rate.
The network in Fig. \ref{fig:anchor_net} is fully convolutional, therefore it can accept input of any height ($H$) and width ($W$), making it versatile to any video resolution.

\begin{figure}
    \centering
    \includegraphics[width=0.7\linewidth]{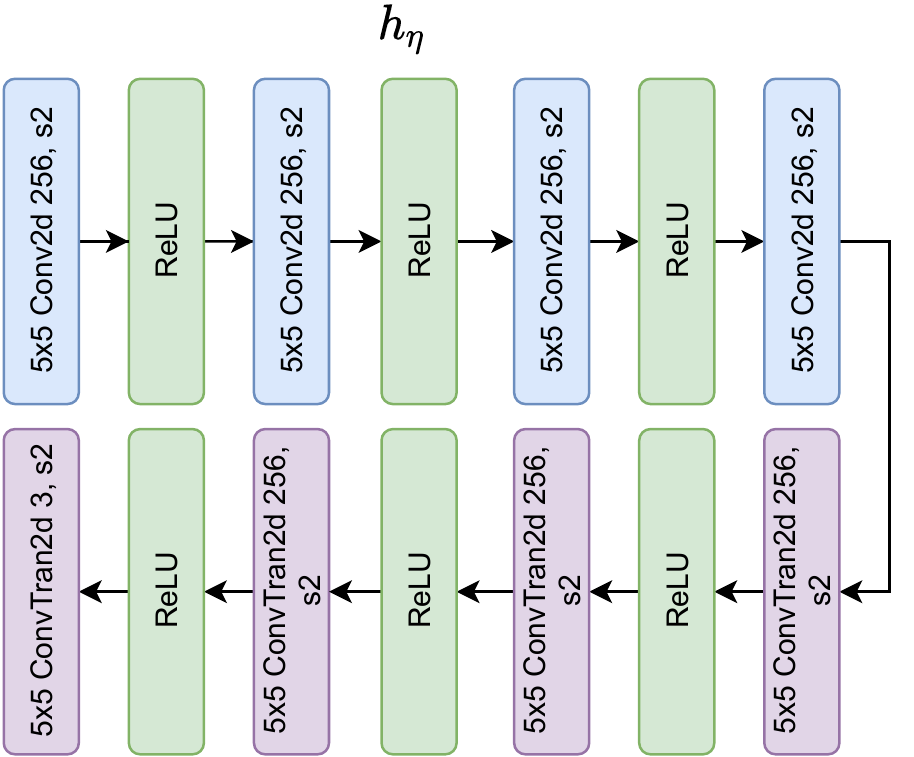}
    \caption{Architecture of the SSF estimator network $h_{\boldsymbol{\eta}}$. 
    }
    \label{fig:ssf_net}
\end{figure}

For the remaining frames, i.e., $\mathbf{x}_i^n$, $i=1,2,\dots,N-1$, we use the interpolation encoder $g_{\boldsymbol{\phi}}(\cdot)$ to encode the motion information ($\mathbf{f}_{i-t}^n$, $\mathbf{f}_{i+t}^n$) and residual information ($\mathbf{r}_{i-t}^n$, $\mathbf{r}_{i+t}^n$) of $\mathbf{x}_i^n$ with respect to two reference frames ($\bar{\mathbf{x}}_{i-t}^n$, $\bar{\mathbf{x}}_{i+t}^n$) that are $t$ frames away from the current frame. 
The reference frames $\bar{\mathbf{x}}_{i-t}^n$, $\bar{\mathbf{x}}_{i+t}^n$ are what the encoder expects the corresponding reconstructed frames $\hat{\mathbf{x}}_{i-t}^n$, $\hat{\mathbf{x}}_{i+t}^n$ to be.
This is done via channel emulation and decoding at the transmitter side to obtain an approximation of what the transmitter expects the receiver to reconstruct.

\begin{figure*}
    \centering
    \includegraphics[width=\textwidth]{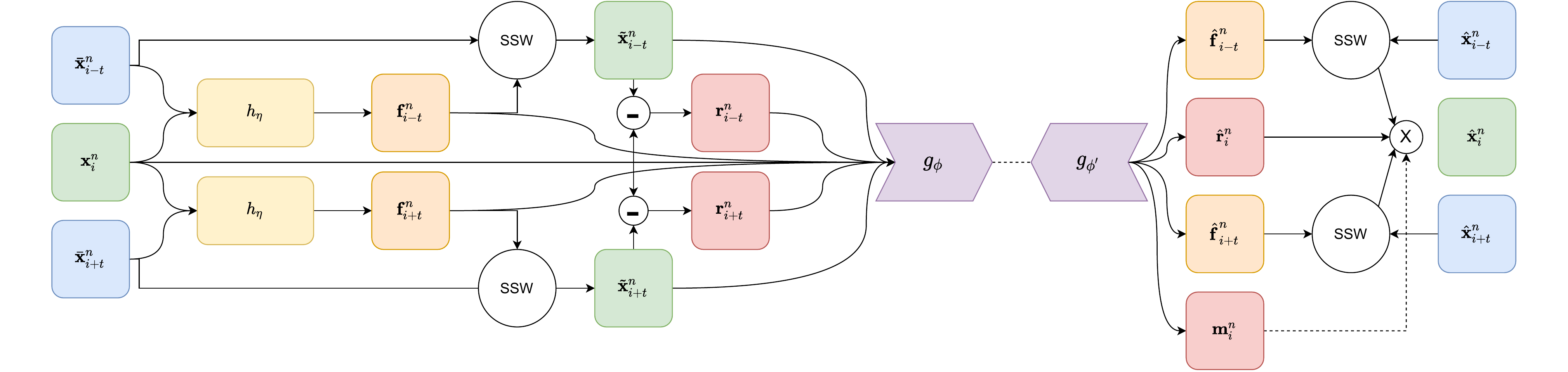}
    \caption{Information flow over the interpolation network.}
    \label{fig:interp_net}
\end{figure*}

We follow the same interpolation structure as the one presented in Fig. \ref{fig:codec_interp_structure} for a GoP size $N=4$.
That is, for $i=2$, $t=2$, while for $i=1,3$, $t=1$.
We define $\bar{\mathbf{x}}_0^n=\bar{\mathbf{x}}_N^{n-1}$, and assume that the GoPs are encoded and decoded sequentially, such that the frames from the previous decoded GoP are available as reference for the current GoP.
To interpolate $\mathbf{x}_i^n$ from $\bar{\mathbf{x}}_{i-t}^n$ and $\bar{\mathbf{x}}_{i+t}^n$, motion information, such as optical flow \cite{beauchemin_computation_1995}, is usually used to warp a reference image by translating the pixels in the reference image according to the optical flow vectors. 
These optical flow vectors describe the horizontal and vertical translations of each pixel in a reference image in order to transform it into the target image.
The difference between the optical flow transformed image and the true target image is called the \textit{residual}, which is used to capture information that cannot be described by optical flow, such as occlusion/disocclusion and camera movements.

To estimate the motion information ($\mathbf{f}_{i-t}^n$, $\mathbf{f}_{i+t}^n$), instead of using optical flow, we use scaled space flow (SSF), which was first proposed by \cite{agustsson_scale-space_2020} as a more general description of pixel warping than optical flow.
The idea is to blur regions of the frame where the motion is difficult to model using traditional pixel warping and instead compensate those regions using the residual.
To that end, in scale-space warping (SSW), a frame is first transformed into a fixed-resolution volume $\bar{\mathbf{X}}_{i+t}^n=[\bar{\mathbf{x}}_{i+t}^n,\bar{\mathbf{x}}_{i+t}^n\otimes G(\sigma_0),\bar{\mathbf{x}}_{i+t}^n\otimes G(2\sigma_0),\dots,\bar{\mathbf{x}}_{i+t}^n\otimes G(2^{V-1}\sigma_0)]$, where $\bar{\mathbf{x}}_{i+t}^n\otimes G(\sigma_0)$ denotes Gaussian blurring of the frame $\bar{\mathbf{x}}_{i+t}^n$ by convolving $\bar{\mathbf{x}}_{i+t}^n$ with a Gaussian kernel $G(\sigma_0)$ with standard deviation $\sigma_0$ and $\otimes$ is the convolution operation.
$\bar{\mathbf{X}}_{i+t}^n\in\mathbb{R}^{H\times W\times (V+1)}$ represents a progressively blurred version of $\mathbf{x}_{i+t}^n$, which can be sampled at continuous points via trilinear interpolation.
The scaled space flow $\mathbf{f}_{i+t}^n\in\mathbb{R}^{H\times W\times 3}$ that warps frame $\bar{\mathbf{x}}_{i+t}^n$ to an approximation of $\mathbf{x}_i^n$ denoted by $\Tilde{\mathbf{x}}_{i+t}^n$ is then defined as 

\begin{align}
   &\Tilde{\mathbf{x}}_{i+t}^n = \text{SSW}(\bar{\mathbf{x}}_{i+t}^n,\mathbf{f}_{i+t}^n)\\
   &\text{s.t.}~\Tilde{\mathbf{x}}_{i+t}^n[x,y]\nonumber \\
   &~~~= \bar{\mathbf{X}}_{i+t}^n[x + \mathbf{f}_{i+t}^n[x,y,1], y + \mathbf{f}_{i+t}^n[x,y,2], \mathbf{f}_{i+t}^n[x,y,3]]
\end{align}
To estimate the scaled space flow $\mathbf{f}_{i+t}^n$, we use the network architecture $h_{\boldsymbol{\eta}}:\mathbb{R}^{H\times W\times 6}\mapsto\mathbb{R}^{H\times W\times 3}$ proposed in \cite{agustsson_scale-space_2020},
\begin{equation}
    \mathbf{f}_{i+t}^n=h_{\boldsymbol{\eta}}(\mathbf{x}_i^n,\hat{\mathbf{x}}_{i+t}^n).
\end{equation}
Fig. \ref{fig:ssf_net} shows the architecture of the SSF estimator network $h_{\boldsymbol{\eta}}$.
This architecture is similar to U-Net \cite{ronneberger_u-net_2015}, which was first proposed for bioimage segmentation.
The architecture progressively downsamples the input using convolutional layers before upsampling it back to the frame dimensions.
The channel dimension of the output SSF $\mathbf{f}_{i+t}^n$, instead of representing the three color channels, represent the $(x,y,z)$ sampling points in the SSF volume $\bar{\mathbf{X}}_{i+t}^n$.

Given the above definition of SSF, the residual $\mathbf{r}_{i+t}^n$ is defined as
\begin{equation}
    \mathbf{r}_{i+t}^n = \mathbf{x}_i^n - \Tilde{\mathbf{x}}_{i+t}^n.
    \label{eq:res}
\end{equation}

The interpolation encoder $g_{\boldsymbol{\phi}}:\mathbb{R}^{H\times W\times 21}\mapsto\mathbb{C}^{k}$ defines the mapping
\begin{align}
   &\mathbf{z}_i^n = g_{\boldsymbol{\phi}}(\mathbf{x}_i^n, \bar{\mathbf{x}}_{i-t}^n, \bar{\mathbf{x}}_{i+t}^n, \mathbf{r}_{i-t}^n, \mathbf{r}_{i+t}^n, \mathbf{f}_{i-t}^n, \mathbf{f}_{i+t}^n, \hat{\sigma}^2), \nonumber\\
   & i=1,2,\dots,N-1.
\end{align}
The vector $\mathbf{z}_i^n\in\mathbb{C}^k$ is power normalized according to Eqn. (\ref{eq:power_norm}) and sent across the channel according to Eqn. (\ref{eq:awgn_channel}).

Given the noisy $\hat{\mathbf{y}}_i^n$, the decoder first estimates the SSF, the residual, and a mask.
That is, the interpolation decoder $g_{\boldsymbol{\phi}^\prime}:\mathbb{C}^k\mapsto\mathbb{R}^{H\times W\times 12}$ defines the mapping
\begin{align}
    (\hat{\mathbf{f}}_{i-t}^n,\hat{\mathbf{f}}_{i+t}^n,\hat{\mathbf{r}}_i^n, \mathbf{m}_i^n) = g_{\boldsymbol{\phi}^\prime}(\hat{\mathbf{y}}_i^n, \hat{\sigma}^2),
\end{align}
where $\hat{\mathbf{f}}_{i\pm t}^n\in\mathbb{R}^{H\times W\times 3}$, $\hat{\mathbf{r}}_i^n\in\mathbb{R}^{H\times W\times 3}$, and $\mathbf{m}_i^n\in\mathbb{R}^{H\times W\times 3}$.
$\mathbf{m}_{i,c}^n\in\mathbb{R}^{H\times W}, c=1,2,3$, a 2D matrix in the third dimension of $\mathbf{m}_i^n$, satisfies:
\begin{equation}
    \sum_{c=1}^3 \mathbf{m}_{i,c}^n=\mathbf{1}_{H\times W}.
\end{equation}
That is, for each $H$ and $W$ index of the mask $\mathbf{m}_i^n$, the sum of values along the channel dimension is equal to $1$, which is achieved by using the softmax activation.
The reconstructed frame is then defined as:
\begin{align}
    \hat{\mathbf{x}}_i^n = &(\mathbf{m}_i^n)_1\ast\text{SSW}(\hat{\mathbf{x}}_{i-t}^n, \hat{\mathbf{f}}_{i-t}^n) + \nonumber\\
    &(\mathbf{m}_i^n)_2\ast\text{SSW}(\hat{\mathbf{x}}_{i+t}^n, \hat{\mathbf{f}}_{i+t}^n) + (\mathbf{m}_i^n)_3\ast\hat{\mathbf{r}}_i^n,
\end{align}
where $\ast$ refers to element-wise multiplication.
The predicted mask $\mathbf{m}_i^n$ acts as a convex set of weights to sum the two motion compensated predictions and the predicted residual, such that the resultant prediction $\hat{\mathbf{x}}^n_i$ remains within $[0, 255]$.
A diagram of the interpolation structure described herein can be seen in Fig. \ref{fig:interp_net}. 
The architectures of $g_{\boldsymbol{\phi}}$ and $g_{\boldsymbol{\phi}^\prime}$ are functionally the same as $f_{\boldsymbol{\theta}}$ and $f_{\boldsymbol{\theta}^\prime}$, except the size of the input tensor, which is the concatenation of $(\mathbf{x}_i^n, \bar{\mathbf{x}}_{i-t}^n, \bar{\mathbf{x}}_{i+t}^n, \mathbf{r}_{i-t}^n, \mathbf{r}_{i+t}^n, \mathbf{f}_{i-t}^n, \mathbf{f}_{i+t}^n)$ along the channel dimension.
As in the key frame, the network weights $(\boldsymbol{\phi},\boldsymbol{\phi}^\prime, \boldsymbol{\eta})$ are updated via backpropagation with respect to the gradient of the loss between the original and the reconstructed frame.

\subsection{Bandwidth Allocation}
\label{subsec:bw_alloc}

In the previous section, we have assumed that each frame utilizes the full bandwidth of $k$ channel uses allowed for each GoP.
In order to satisfy the bandwidth constraint defined in Section \ref{sec:problem_form}, the encoder must decide how to allocate $k$ channel uses to the $N$ frames in a GoP.
Intuitively, if the frame in consideration $\mathbf{x}_i^n$ is exactly the same with respect to the reference frames $(\bar{\mathbf{x}}_{i-t}^n,\bar{\mathbf{x}}_{i+t}^n)$ that it is interpolated from, then no information needs to be transmitted.
On the other hand, if there is significant differences with respect to the reference frames, then more information needs to be sent in order to accurately interpolate the frame.
Since the last frame of a previous GoP becomes the reference frame of the next GoP ($\bar{\mathbf{x}}_{N}^n=\bar{\mathbf{x}}_{0}^{n+1}$), we formulate the problem of allocating available bandwidth in each GoP as a Markov decision process (MDP) and solve the optimal bandwidth allocation policy using reinforcement learning. 

A MDP is defined by the tuple $(\mathcal{S},\mathcal{A},P,r)$, where $\mathcal{S}$ is the set of states, $\mathcal{A}$ is the action set, $P$ is the probability transition kernel that defines the probability of one state transitioning to another state given an action, and $r:\mathcal{S}\times\mathcal{A}\mapsto\mathbb{R}$ is the reward function.
At each time step $n$, an agent observes state $\mathbf{s}^n\in\mathcal{S}$ and takes an action $\mathbf{a}^n\in\mathcal{A}$ based on its policy $\pi:\mathcal{S}\mapsto\mathcal{A}$. 
The state then transitions to $\mathbf{s}^{n+1}$ according to the probability $P(\mathbf{s}^{n+1}|\mathbf{s}^{n},\mathbf{a}^{n})$, and the agent receives a reward $r^n(\mathbf{s}^n,\mathbf{a}^n)$.
The objective is to maximize the expected sum of rewards $J(\pi)=\mathbb{E}_{\mathbf{s}_1\sim\omega_{\mathbf{s}_1},\pi}[\sum_{i=1}^\infty\gamma^{(i-1)}r^i(\mathbf{s}^i,\mathbf{a}^i)]$, where $\omega_{\mathbf{s}_1}$ is the initial state distribution and $\gamma\in(0,1)$ is the reward discount factor to ensure convergence.

In the dynamic bandwidth allocation problem, we define the state at time step $n$ as $\mathbf{s}^n=\{\mathbf{M}^n,\mathbf{R}^n,\mathbf{F}^n,\hat{\sigma}^2\}$, where
\begin{align}
    &\mathbf{M}^n=\{\bar{\mathbf{x}}_i^{n}\}_{i=0}^{N}, \\
    &\mathbf{R}^n=\{\mathbf{r}_{i\pm t}^n\}_{i=1}^{N-1},\\
    &\mathbf{F}^n=\{\mathbf{f}_{i\pm t}^n\}_{i=1}^{N-1},
\end{align}
where $\bar{\mathbf{x}}_0^{n}=\bar{\mathbf{x}}_N^{n-1}$.
The action set $\mathcal{A}$ is the set of all the different ways the available bandwidth $k$ can be allocated to each frame in the GoP.
In order for the decoder functions $f_{\boldsymbol{\theta}^\prime}$ and $g_{\boldsymbol{\phi}^\prime}$ to be able to decode each frame that has been given different amounts of bandwidth, we use a result in \cite{kurka_bandwidth-agile_2021}, which showed that joint source-channel encoded images can be successively refined by sending increasingly more information.
This is achieved by dividing the latent vectors $\mathbf{z}_i^n$ into $V$ equal sized blocks (i.e., $\mathbf{z}_i^n=\{\mathbf{z}_{i,1}^n,\dots,\mathbf{z}_{i,V}^n\},~\mathbf{z}_{i,v}^n\in\mathbb{C}^{\frac{k}{V}}, v=1,...,V$), while randomly varying the number of blocks $v_i^n$ of the latent code transmitted in each batch $\mathbf{z}_{i}^n(v_i^n)=\{\mathbf{z}_{i,1}^n,\dots,\mathbf{z}_{i,v_i^n}^n\},~v_i^n\leq V$.
This training process leads to the descending ordering of information from $\mathbf{z}_{i,1}^n$ to $\mathbf{z}_{i,V}^n$.
As such, each action represents $\mathbf{a}^n=[v_1^n,...,v_N^n]$.
We implement this training process in the algorithm described in Section \ref{subsec:video_jscc} by zeroing out the blocks in the latent vector not transmitted.
As such, the action set is all the ways to assign $V$ blocks to the $N$ frames in the GoP; that is, $\sum_{i=1}^N v_i^n = k$.
Consequently, it can be shown that the number of ways to assign $V$ blocks to $N$ sets without replacement is
\begin{equation}
    |\mathcal{A}|=\frac{(V+N-1)!}{V!(N-1)!}.
\end{equation}

\begin{figure}
    \centering
    \includegraphics[width=\linewidth]{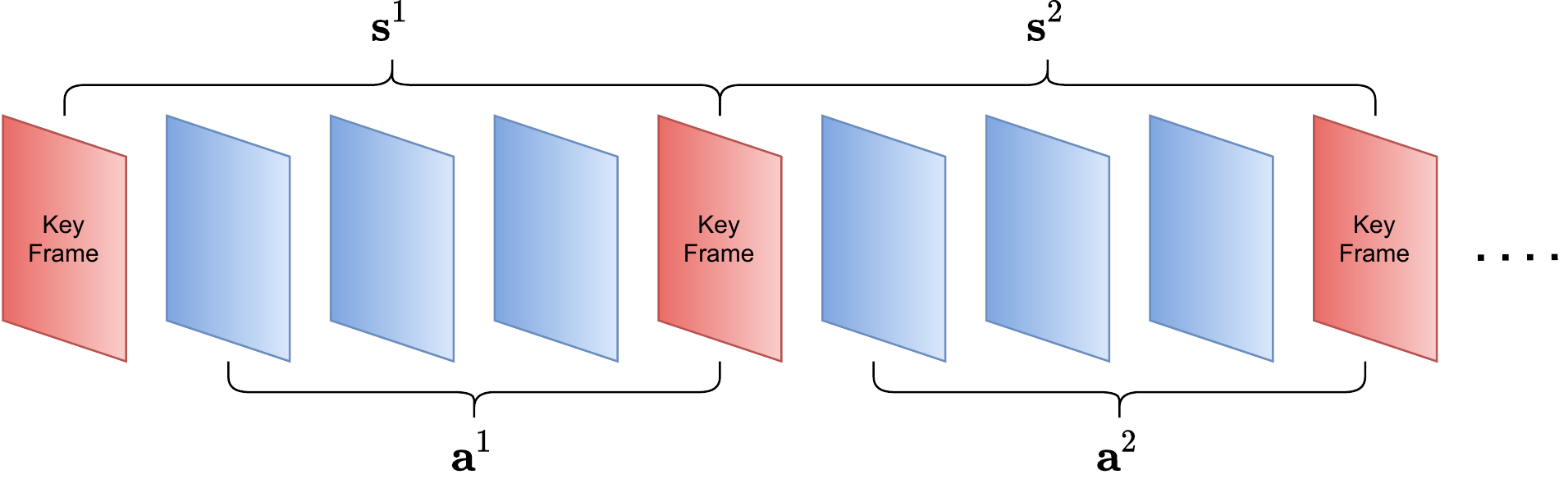}
    \caption{Diagram illustrating the bandwidth allocation problem formulated as a MDP. 
    }
    \label{fig:allocation_structure}
\end{figure}

Since we are concerned with maximizing the visual quality of the final video, we define the reward function $r^n$ as
\begin{align}
    &r^n=-\log_{10}\Big(l(\mathbf{X}^n,\hat{\mathbf{X}}^n)\Big),
\end{align}
where $l(\cdot, \cdot)$ is either $l_{\text{PSNR}}$ or $l_{\text{MS-SSIM}}$ depending on the video quality metric used.
Note that in the previous section, we said that the transmitter performs channel emulation in order to obtain the reference frames ($\bar{\mathbf{x}}_{i-t}^n,\bar{\mathbf{x}}_{i+t}^n$).
Since the precise estimate of the reference frames is bootstrapped to the amount of bandwidth allocated, we initially assume a uniform bandwidth allocation (i.e., $v_i^n=V/N$) when computing the SSF and residuals; but once the allocation has been done, the reference frame in the next state (i.e., $\bar{\mathbf{x}}_0^{n+1}\in\mathbf{M}^{n+1}$) is estimated based on the bandwidth allocated.

To solve the MDP described herein and to learn the optimal allocation policy, we use deep Q-learning \cite{mnih_human-level_2015}, where the network $q_{\boldsymbol{\psi}}$ seeks to approximate the Q-function $Q:\mathcal{S}\times\mathcal{A}\rightarrow\mathbb{R}$.
The purpose of the Q-function is to map each state and action pair to a Q value, which represents the total discounted reward from step $n$ given the state and action pair $(\mathbf{s}^n,\mathbf{a}^n)$. 
That is,
\begin{equation}
    Q(\mathbf{s}^n,\mathbf{a}^n)=E\Bigg[\sum_{i=n}^\infty \gamma^{i-n}r^i\Bigg|\mathbf{s}^n,\mathbf{a}^n\Bigg],~\forall(\mathbf{s}^n,\mathbf{a}^n)\in\mathcal{S}\times\mathcal{A}.
\end{equation}
As is typical in DQN methods, we use \textit{replay buffer}, \textit{target network}, and \textit{$\epsilon$-greedy} to aid the learning of the Q-function.
The replay buffer $\mathcal{R}$ stores experiences $(\mathbf{s}^n,\mathbf{a}^n,r^n,\mathbf{s}^{n+1})$ and are sampled uniformly to update the parameters $\boldsymbol{\psi}$.
This prevents the states from being correlated, which would break the assumption in most optimization algorithms that the samples are independent. 
We use target parameters ${\boldsymbol{\psi}^-}$, which are copies of ${\boldsymbol{\psi}}$, to compute the DQN loss function:
\begin{align}
    L_{\text{DQN}}(\boldsymbol{\psi})=\Big(r^n&+\gamma\max_{\mathbf{a}}\big\{q_{\boldsymbol{\psi}^-}\big(\mathbf{s}^{n+1},\mathbf{a}\big)\big\} - q_{\boldsymbol{\psi}} \big(\mathbf{s}^n,\mathbf{a}^n\big)\Big)^2.
    \label{eq:dqn_loss}
\end{align}
The parameters $\boldsymbol{\psi}$ are then updated via gradient descent according to the gradient $\nabla_{\boldsymbol{\psi}}L_{\text{DQN}}(\boldsymbol{\psi})$. 
In practice, the DQN loss is approximated with a batch of samples from the replay buffer $\mathcal{B}\subset\mathcal{R}$. 

\begin{figure}
    \centering
    \includegraphics[width=\linewidth]{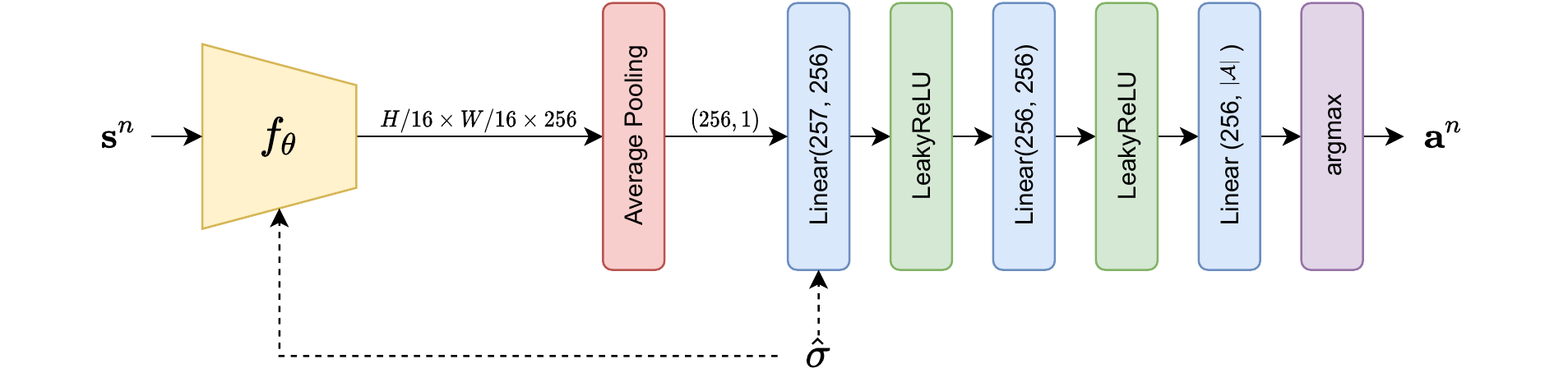}
    \caption{Architecture of the bandwidth allocation network $q_{\boldsymbol{\psi}}$. 
    The convolutional part of the network for feature extraction is functionally the same as the key encoder network ($f_{\boldsymbol{\theta}}$) but with $21(N-1)+6$ input dimensions to account for all the tensors in the state $\mathbf{s}^n$.}
    \label{fig:bwagent_net}
\end{figure}

The target network parameters are updated via
\begin{equation}
    \boldsymbol{\psi}^-\leftarrow\tau\boldsymbol{\psi}+(1-\tau)\boldsymbol{\psi}^-,
    \label{eq:target_update}
\end{equation}
where $0\leq\tau\leq1$.
The target networks here stabilize the updates. 
Due to Q-learning being bootstrapped, if the same $q_{\boldsymbol{\psi}}$ is used to estimate the state-action value of GoP number $n$ and $n+1$, both values would move at the same time, which may lead to the updates to never converge.
By introducing the target networks, this effect is reduced due to the much slower updates of the target network, as done in Eqn. (\ref{eq:target_update}).

To promote exploration, we use $\epsilon$-greedy, which chooses a random action with probability $\epsilon$ at each GoP. 
That is,
\begin{equation}
    \textbf{a}_n=\begin{cases}
               \argmax_{\mathbf{a}}q_{\boldsymbol{\psi}}(\textbf{s}_n,\mathbf{a}),~&\text{w.p. }1-\epsilon\\
               \mathbf{a} \sim \text{Uniform}(\mathcal{A}),~&\text{w.p. }\epsilon
            \end{cases},
\end{equation}
where $\mathbf{a} \sim \text{Uniform}(\mathcal{A})$ denotes an action that is sampled uniformly from the action set $\mathcal{A}$.
A diagram of the architecture used for $q_{\boldsymbol{\psi}}$ is shown in Fig. \ref{fig:bwagent_net}.

Upon initialization, we send the first frame $\mathbf{x}_1^1$ using full bandwidth $k$.
The first frame can be considered as a GoP on its own. 
For all subsequent GoPs, we perform optimal bandwidth allocation as described in this section.
A diagram illustrating the bandwidth allocation problem formulated herein is shown in Fig. \ref{fig:allocation_structure}.

\section{Numerical Results}
\label{sec:results}

\subsection{Training Details}
\label{subsec:training}

We train our models on the UCF101 dataset \cite{soomro_ucf101_2012} using Pytorch \cite{paszke_automatic_2017}, with the Adam optimizer \cite{kingma_adam_2017} at learning rate $1e^{-5}$. 
We split the dataset with $0.8:0.1:0.1$ ratio between training, validation and evaluation. 
We train the JSCC ($f_{{\boldsymbol{\theta}}}, f_{\boldsymbol{\theta}^\prime}, g_{\boldsymbol{\phi}}, g_{\boldsymbol{\phi}^\prime}, h_{\boldsymbol{\eta}}$) networks first, randomizing the latent vector block sizes as described in Section \ref{subsec:bw_alloc}, before we train the bandwidth allocation network $q_{\boldsymbol{\psi}}$ to find the optimal bandwidth allocation policy.
We also assume that during this phase, the transmitter and receiver can estimate the channel SNR accurately $\hat{\sigma}^2=\sigma^2$.
We use a training batch size of $4$ when training the JSCC networks and a batch size of $8$ when training the bandwidth allocation network.
The batch sizes are relatively small due to memory restrictions of the available GPU (11GB Nvidia RTX 2080Ti).
We use early stopping based on the validation error with a patience of 8 epochs.
We adjust the learning rate based on the number of bad epochs: if the validation error does not improve for 4 epochs in a row, the learning rate is multiplied by $0.8$.

We define the SNR estimated by the transmitter and receiver to be
\begin{equation}
    \text{SNR}_{\text{Est}} = 10\log_{10}\Big(\frac{P}{\hat{\sigma}^2}\Big).
\end{equation}

For training the bandwidth allocation network, we choose DQN hyper-parameters $\gamma=0.99$, $\tau=0.005$, and a replay buffer size $|\mathcal{R}|=1000$.
The function used for $\epsilon$-greedy exploration is
\begin{equation}
    \epsilon=\epsilon_{\text{end}}+(\epsilon_0-\epsilon_{\text{end}})e^{\big(-\frac{\text{episode}}{\lambda}\big)},
\end{equation}
where $\lambda$ controls the decay rate of $\epsilon$, and in each episode the bandwidth allocation network allocates the bandwidth resource within one video sequence.
We choose $\epsilon_0=0.9$, $\epsilon_{\text{end}}=0.05$, and $\lambda=1000$ for these parameters.

We train our model at different channel SNRs and evaluate each model at the same range of SNRs.
In each batch, the training SNR is sampled uniformly from the range $[-5,20]$dB.
We chose $N=4$ and $V=20$ to train our models.
Note that although we can choose $V=k$ to achieve very fine grained bandwidth control, this would lead to a very large action space, which makes Q-learning difficult.
We let $P=1$ and compute the necessary $\sigma^2$ to achieve the desired SNR.

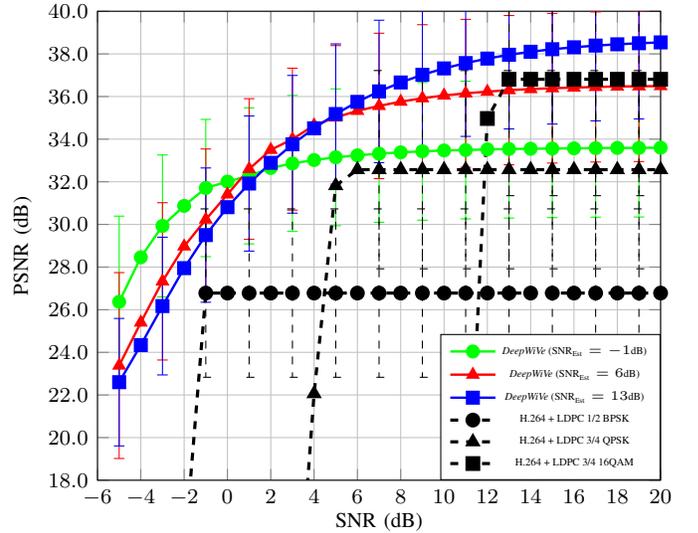
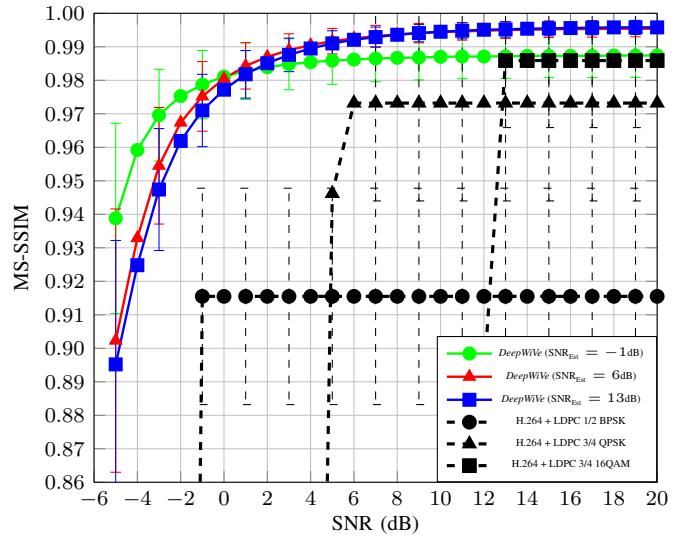
\begin{figure} 
    \centering
  \subfloat[PSNR\label{subfig:psnr_graceful_h264_kn0031}]{%
    \begin{tikzpicture}
        \pgfplotsset{
            legend style={
                font=\fontsize{4}{4}\selectfont,
                at={(1.0,.0)},
                anchor=south east,
            },
            width=0.5\textwidth,
            xmin=-6,
            xmax=20,
            ymin=18,
            ymax=40,
            xtick distance=2,
            ytick distance=2,
            xlabel={SNR (dB)},
            ylabel={PSNR (dB)},
            grid=both,
            grid style={line width=.1pt, draw=gray!10},
            major grid style={line width=.2pt,draw=gray!50},
            every axis/.append style={
                x label style={
                    font=\fontsize{8}{8}\selectfont,
                    at={(axis description cs:0.5,-0.05)},
                    },
                y label style={
                    font=\fontsize{8}{8}\selectfont,
                    at={(axis description cs:-0.1,0.5)},
                    },
                x tick label style={
                    font=\fontsize{8}{8}\selectfont,
                    /pgf/number format/.cd,
                    fixed,
                    fixed zerofill,
                    precision=0,
                    /tikz/.cd
                    },
                y tick label style={
                    font=\fontsize{8}{8}\selectfont,
                    /pgf/number format/.cd,
                    fixed,
                    fixed zerofill,
                    precision=1,
                    /tikz/.cd
                    },
            }
        }
        \begin{axis}
        \addplot[green, solid, line width=0.9pt, mark=*, mark options={fill=green, scale=1.1}, error bars/.cd, y dir=both, y explicit, every nth mark=2] 
                table [x=snrtest, y=PSNR(snrtrain-1), y error=PSNR_error(snrtrain-1), col sep=comma] {Data/deepwive_graceful_psnr_c240_tsnr-1.csv};
        \addlegendentry{\textit{DeepWiVe} ($\text{SNR}_{\text{Est}}=-1$dB)}
        
        \addplot[red, solid, line width=0.9pt, mark=triangle*, mark options={fill=red, scale=1.1}, error bars/.cd, y dir=both, y explicit, every nth mark=2] 
                table [x=snrtest, y=PSNR(snrtrain6), y error=PSNR_error(snrtrain6), col sep=comma]
                {Data/deepwive_graceful_psnr_c240_tsnr6.csv};
        \addlegendentry{\textit{DeepWiVe} ($\text{SNR}_{\text{Est}}=6$dB)}
        
        \addplot[blue, solid, line width=0.9pt, mark=square*, mark options={fill=blue, scale=1.1, solid}, error bars/.cd, y dir=both, y explicit, every nth mark=2] 
                table [x=snrtest, y=PSNR(snrtrain13), y error=PSNR_error(snrtrain13), col sep=comma] {Data/deepwive_graceful_psnr_c240_tsnr13.csv};
        \addlegendentry{\textit{DeepWiVe} ($\text{SNR}_{\text{Est}}=13$dB)}
        
        \addplot[color=black, dashed, line width=1.2pt, mark=*, mark options={fill=black, solid, scale=1.1}, error bars/.cd, y dir=both, y explicit, every nth mark=2] 
        table [x=snrtest, y=bpsk, y error=bpsk_error, col sep=comma]{Data/h264_1,2LDPC_psnr_c240.csv};
        \addlegendentry{H.264 + LDPC 1/2 BPSK}
        
        \addplot[color=black, dashed, line width=1.2pt, mark=triangle*, mark options={fill=black, solid, scale=1.1}, error bars/.cd, y dir=both, y explicit, every nth mark=2] 
        table [x=snrtest, y=qpsk, y error=qpsk_error, col sep=comma]{Data/h264_2,3LDPC_psnr_c240.csv};
        \addlegendentry{H.264 + LDPC 3/4 QPSK}
        
        \addplot[color=black, dashed, line width=1.2pt, mark=square*, mark options={fill=black, solid, scale=1.1}, error bars/.cd, y dir=both, y explicit, every nth mark=2] 
        table [x=snrtest, y=16qam, y error=16qam_error, col sep=comma]{Data/h264_2,3LDPC_psnr_c240.csv};
        \addlegendentry{H.264 + LDPC 3/4 16QAM}
        
        \end{axis}
        \end{tikzpicture}
    }
    \\
  \subfloat[MS-SSIM\label{subfig:msssim_graceful_h264_kn0031}]{%
    \begin{tikzpicture}
        \pgfplotsset{
            legend style={
                font=\fontsize{4}{4}\selectfont,
                at={(1.0,0.)},
                anchor=south east,
            },
            width=0.5\textwidth,
            xmin=-6,
            xmax=20,
            ymin=0.86,
            ymax=1,
            xtick distance=2,
            ytick distance=0.01,
            xlabel={SNR (dB)},
            ylabel={MS-SSIM},
            grid=both,
            grid style={line width=.1pt, draw=gray!10},
            major grid style={line width=.2pt,draw=gray!50},
            every axis/.append style={
                x label style={
                    font=\fontsize{8}{8}\selectfont,
                    at={(axis description cs:0.5,-0.05)},
                    },
                y label style={
                    font=\fontsize{8}{8}\selectfont,
                    at={(axis description cs:-0.1,0.5)},
                    },
                x tick label style={
                    font=\fontsize{8}{8}\selectfont,
                    /pgf/number format/.cd,
                    fixed,
                    fixed zerofill,
                    precision=0,
                    /tikz/.cd
                    },
                y tick label style={
                    font=\fontsize{8}{8}\selectfont,
                    /pgf/number format/.cd,
                    fixed,
                    fixed zerofill,
                    precision=2,
                    /tikz/.cd
                    },
            }
        }
        \begin{axis}
        \addplot[green, solid, line width=0.9pt, mark=*, mark options={fill=green, scale=1.1}, error bars/.cd, y dir=both, y explicit, every nth mark=2] 
                table [x=snrtest, y=MS-SSIM(snrtrain-1), y error=MS-SSIM_error(snrtrain-1), col sep=comma] {Data/deepwive_graceful_msssim_c240_tsnr-1.csv};
        \addlegendentry{\textit{DeepWiVe} ($\text{SNR}_{\text{Est}}=-1$dB)}
        
        \addplot[red, solid, line width=0.9pt, mark=triangle*, mark options={fill=red, scale=1.1}, error bars/.cd, y dir=both, y explicit, every nth mark=2] 
                table [x=snrtest, y=MS-SSIM(snrtrain6), y error=MS-SSIM_error(snrtrain6), col sep=comma]
                {Data/deepwive_graceful_msssim_c240_tsnr6.csv};
        \addlegendentry{\textit{DeepWiVe} ($\text{SNR}_{\text{Est}}=6$dB)}
        
        \addplot[blue, solid, line width=0.9pt, mark=square*, mark options={fill=blue, scale=1.1, solid}, error bars/.cd, y dir=both, y explicit, every nth mark=2] 
                table [x=snrtest, y=MS-SSIM(snrtrain13), y error=MS-SSIM_error(snrtrain13), col sep=comma] {Data/deepwive_graceful_msssim_c240_tsnr13.csv};
        \addlegendentry{\textit{DeepWiVe} ($\text{SNR}_{\text{Est}}=13$dB)}
        
        \addplot[color=black, dashed, line width=1.2pt, mark=*, mark options={fill=black, solid, scale=1.1}, error bars/.cd, y dir=both, y explicit, every nth mark=2] 
        table [x=snrtest, y=bpsk, y error=bpsk_error, col sep=comma] {Data/h264_1,2LDPC_msssim_c240.csv};
        \addlegendentry{H.264 + LDPC 1/2 BPSK}
        
        \addplot[color=black, dashed, line width=1.2pt, mark=triangle*, mark options={fill=black, solid, scale=1.1}, error bars/.cd, y dir=both, y explicit, every nth mark=2] 
        table [x=snrtest, y=qpsk, y error=qpsk_error, col sep=comma] {Data/h264_2,3LDPC_msssim_c240.csv};
        \addlegendentry{H.264 + LDPC 3/4 QPSK}
        
        \addplot[color=black, dashed, line width=1.2pt, mark=square*, mark options={fill=black, solid, scale=1.1}, error bars/.cd, y dir=both, y explicit, every nth mark=2] 
        table [x=snrtest, y=16qam, y error=16qam_error, col sep=comma] {Data/h264_2,3LDPC_msssim_c240.csv};
        \addlegendentry{H.264 + LDPC 3/4 16QAM}
        \end{axis}
        \end{tikzpicture}
        }
  \caption{Comparison of DeepWiVe using different $\text{SNR}_{\text{Est}}$ to H.264 paired with LDPC codes ($\rho=0.031$).
  }
  \label{fig:graceful_h264_kn0031} 
\end{figure}

\begin{figure*}
    \centering
    \includegraphics[width=\textwidth]{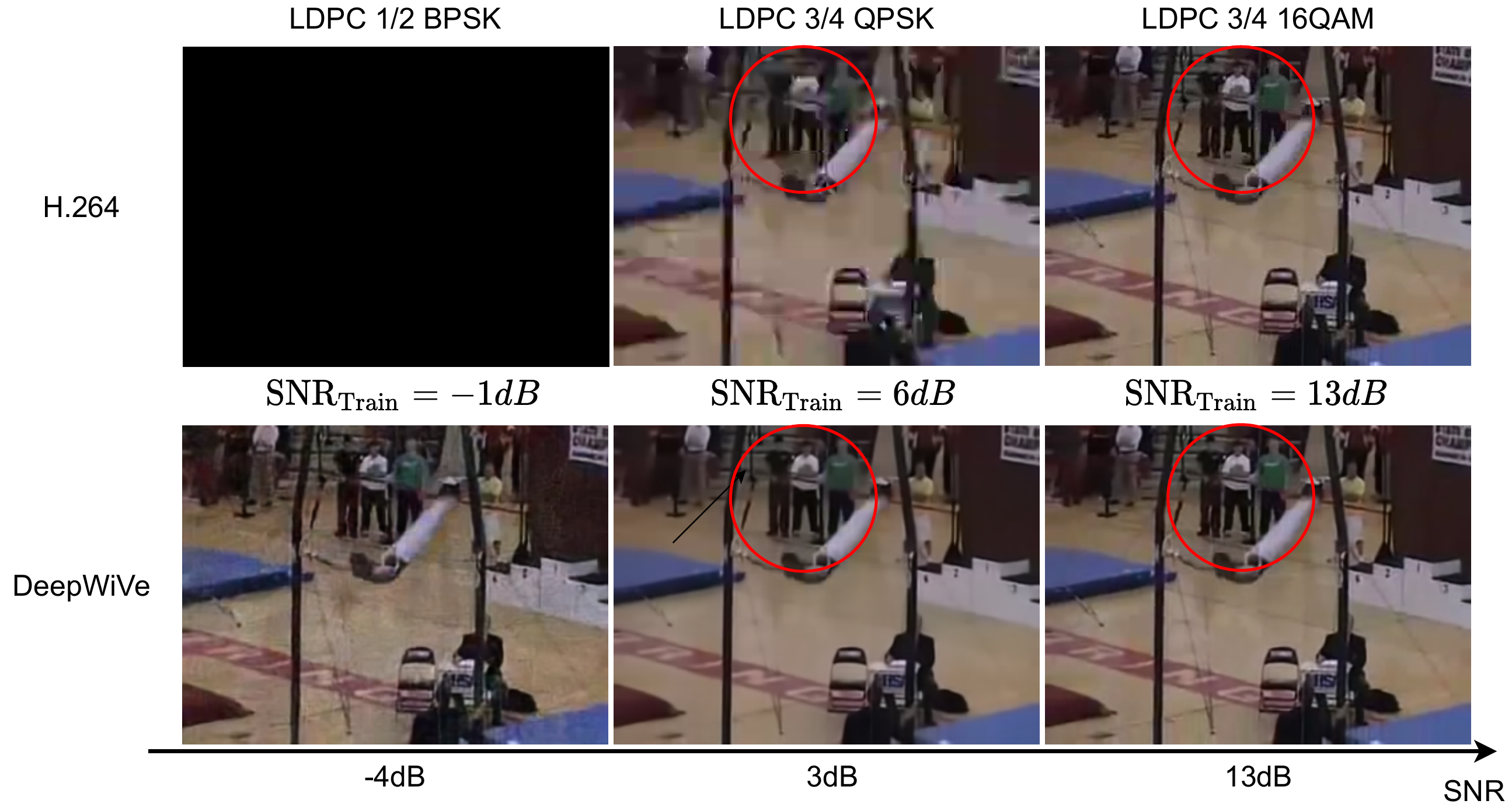}
    \caption{Visual examples of \textit{DeepWiVe} v.s. H.264.
    The difference in video quality can be seen most clearly in the 3 people standing at the back of the scene (encircled).}
    \label{fig:example_frames}
\end{figure*}

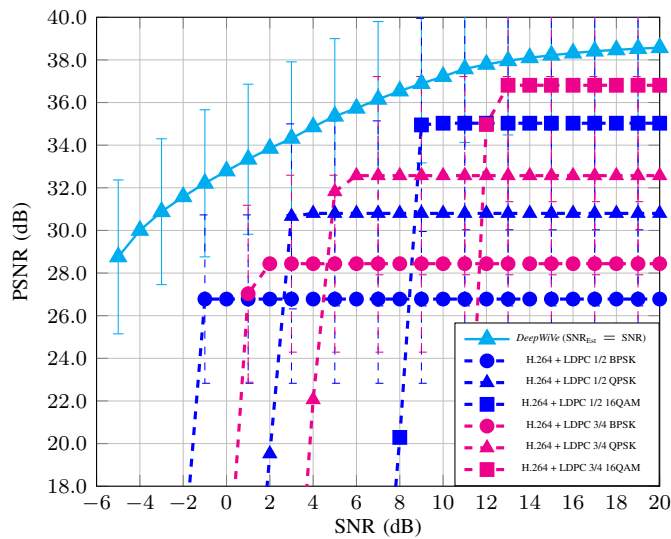
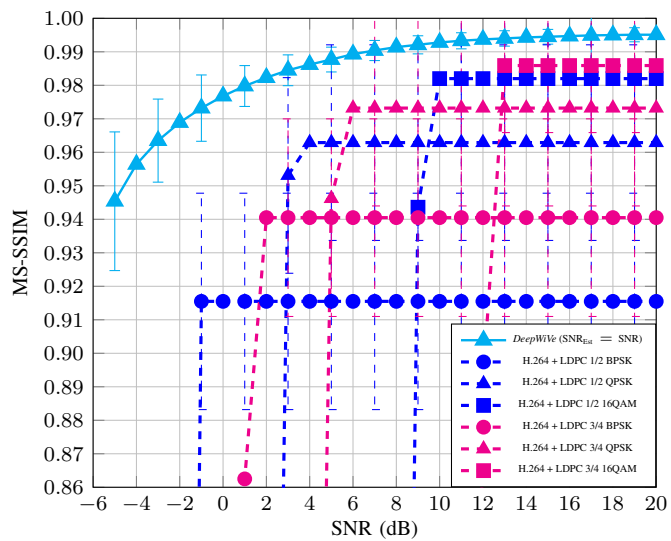
\begin{figure} 
    \centering
  \subfloat[PSNR\label{subfig:psnr_optimal_h264_kn0031}]{%
    \begin{tikzpicture}
        \pgfplotsset{
            legend style={
                font=\fontsize{4}{4}\selectfont,
                at={(1.0,.0)},
                anchor=south east,
            },
            width=0.5\textwidth,
            xmin=-6,
            xmax=20,
            ymin=18,
            ymax=40,
            xtick distance=2,
            ytick distance=2,
            xlabel={SNR (dB)},
            ylabel={PSNR (dB)},
            grid=both,
            grid style={line width=.1pt, draw=gray!10},
            major grid style={line width=.2pt,draw=gray!50},
            every axis/.append style={
                x label style={
                    font=\fontsize{8}{8}\selectfont,
                    at={(axis description cs:0.5,-0.05)},
                    },
                y label style={
                    font=\fontsize{8}{8}\selectfont,
                    at={(axis description cs:-0.1,0.5)},
                    },
                x tick label style={
                    font=\fontsize{8}{8}\selectfont,
                    /pgf/number format/.cd,
                    fixed,
                    fixed zerofill,
                    precision=0,
                    /tikz/.cd
                    },
                y tick label style={
                    font=\fontsize{8}{8}\selectfont,
                    /pgf/number format/.cd,
                    fixed,
                    fixed zerofill,
                    precision=1,
                    /tikz/.cd
                    },
            }
        }
        \begin{axis}
        \addplot[cyan, solid, line width=0.9pt, mark=triangle*, mark options={fill=cyan, scale=1.6}, error bars/.cd, y dir=both, y explicit, every nth mark=2] 
                table [x=snrtest, y=PSNR, y error=PSNR_error, col sep=comma] {Data/deepwive_optimal_c240.csv};
        \addlegendentry{\textit{DeepWiVe} 
        ($\text{SNR}_{\text{Est}}=\text{SNR}$)}
        
        \addplot[color=blue, dashed, line width=1.2pt, mark=*, mark options={fill=blue, solid, scale=1.1}, error bars/.cd, y dir=both, y explicit, every nth mark=2] 
        table [x=snrtest, y=bpsk, y error=bpsk_error, col sep=comma] {Data/h264_1,2LDPC_psnr_c240.csv};
        \addlegendentry{H.264 + LDPC 1/2 BPSK}
        
        \addplot[color=blue, dashed, line width=1.2pt, mark=triangle*, mark options={fill=blue, solid, scale=1.1}, error bars/.cd, y dir=both, y explicit, every nth mark=2] 
        table [x=snrtest, y=qpsk, y error=qpsk_error, col sep=comma] {Data/h264_1,2LDPC_psnr_c240.csv};
        \addlegendentry{H.264 + LDPC 1/2 QPSK}
        
        \addplot[color=blue, dashed, line width=1.2pt, mark=square*, mark options={fill=blue, solid, scale=1.1}, error bars/.cd, y dir=both, y explicit, every nth mark=2] 
        table [x=snrtest, y=16qam, y error=16qam_error, col sep=comma] {Data/h264_1,2LDPC_psnr_c240.csv};
        \addlegendentry{H.264 + LDPC 1/2 16QAM}
        
        
        \addplot[color=magenta, dashed, line width=1.2pt, mark=*, mark options={fill=magenta, solid, scale=1.1}, error bars/.cd, y dir=both, y explicit, every nth mark=2] 
        table [x=snrtest, y=bpsk, y error=bpsk_error, col sep=comma] {Data/h264_2,3LDPC_psnr_c240.csv};
        \addlegendentry{H.264 + LDPC 3/4 BPSK}
        
        \addplot[color=magenta, dashed, line width=1.2pt, mark=triangle*, mark options={fill=magenta, solid, scale=1.1}, error bars/.cd, y dir=both, y explicit, every nth mark=2] 
        table [x=snrtest, y=qpsk, y error=qpsk_error, col sep=comma] {Data/h264_2,3LDPC_psnr_c240.csv};
        \addlegendentry{H.264 + LDPC 3/4 QPSK}
        
        \addplot[color=magenta, dashed, line width=1.2pt, mark=square*, mark options={fill=magenta, solid, scale=1.1}, error bars/.cd, y dir=both, y explicit, every nth mark=2] 
        table [x=snrtest, y=16qam, y error=16qam_error, col sep=comma] {Data/h264_2,3LDPC_psnr_c240.csv};
        \addlegendentry{H.264 + LDPC 3/4 16QAM}
        
        \end{axis}
        \end{tikzpicture}
    }
    \\
  \subfloat[MS-SSIM\label{subfig:msssim_optimal_h264_kn0031}]{%
    \begin{tikzpicture}
        \pgfplotsset{
            legend style={
                font=\fontsize{4}{4}\selectfont,
                at={(1.0,0.)},
                anchor=south east,
            },
            width=0.5\textwidth,
            xmin=-6,
            xmax=20,
            ymin=0.86,
            ymax=1,
            xtick distance=2,
            ytick distance=0.01,
            xlabel={SNR (dB)},
            ylabel={MS-SSIM},
            grid=both,
            grid style={line width=.1pt, draw=gray!10},
            major grid style={line width=.2pt,draw=gray!50},
            every axis/.append style={
                x label style={
                    font=\fontsize{8}{8}\selectfont,
                    at={(axis description cs:0.5,-0.05)},
                    },
                y label style={
                    font=\fontsize{8}{8}\selectfont,
                    at={(axis description cs:-0.1,0.5)},
                    },
                x tick label style={
                    font=\fontsize{8}{8}\selectfont,
                    /pgf/number format/.cd,
                    fixed,
                    fixed zerofill,
                    precision=0,
                    /tikz/.cd
                    },
                y tick label style={
                    font=\fontsize{8}{8}\selectfont,
                    /pgf/number format/.cd,
                    fixed,
                    fixed zerofill,
                    precision=2,
                    /tikz/.cd
                    },
            }
        }
        \begin{axis}[mark options={solid}]
        \addplot[cyan, solid, line width=0.9pt, mark=triangle*, mark options={fill=cyan, scale=1.6}, error bars/.cd, y dir=both, y explicit, every nth mark=2] 
                table [x=snrtest, y=MS-SSIM, y error=MS-SSIM_error, col sep=comma] {Data/deepwive_optimal_c240.csv};
        \addlegendentry{\textit{DeepWiVe} ($\text{SNR}_{\text{Est}}=\text{SNR}$)}
        
        \addplot[color=blue, dashed, line width=1.2pt, mark=*, mark options={fill=blue, solid, scale=1.1}, error bars/.cd, y dir=both, y explicit, every nth mark=2] 
        table [x=snrtest, y=bpsk, y error=bpsk_error, col sep=comma] {Data/h264_1,2LDPC_msssim_c240.csv};
        \addlegendentry{H.264 + LDPC 1/2 BPSK}
        
        \addplot[color=blue, dashed, line width=1.2pt, mark=triangle*, mark options={fill=blue, solid, scale=1.1}, error bars/.cd, y dir=both, y explicit, every nth mark=2] 
        table [x=snrtest, y=qpsk, y error=qpsk_error, col sep=comma] {Data/h264_1,2LDPC_msssim_c240.csv};
        \addlegendentry{H.264 + LDPC 1/2 QPSK}
        
        \addplot[color=blue, dashed, line width=1.2pt, mark=square*, mark options={fill=blue, solid, scale=1.1}, error bars/.cd, y dir=both, y explicit, every nth mark=2] 
        table [x=snrtest, y=16qam, y error=16qam_error, col sep=comma] {Data/h264_1,2LDPC_msssim_c240.csv};
        \addlegendentry{H.264 + LDPC 1/2 16QAM}
        
        
        \addplot[color=magenta, dashed, line width=1.2pt, mark=*, mark options={fill=magenta, solid, scale=1.1}, error bars/.cd, y dir=both, y explicit, every nth mark=2] 
        table [x=snrtest, y=bpsk, y error=bpsk_error, col sep=comma] {Data/h264_2,3LDPC_msssim_c240.csv};
        \addlegendentry{H.264 + LDPC 3/4 BPSK}
        
        \addplot[color=magenta, dashed, line width=1.2pt, mark=triangle*, mark options={fill=magenta, solid, scale=1.1}, error bars/.cd, y dir=both, y explicit, every nth mark=2] 
        table [x=snrtest, y=qpsk, y error=qpsk_error, col sep=comma] {Data/h264_2,3LDPC_msssim_c240.csv};
        \addlegendentry{H.264 + LDPC 3/4 QPSK}
        
        \addplot[color=magenta, dashed, line width=1.2pt, mark=square*, mark options={fill=magenta, solid, scale=1.1}, error bars/.cd, y dir=both, y explicit, every nth mark=2] 
        table [x=snrtest, y=16qam, y error=16qam_error, col sep=comma] {Data/h264_2,3LDPC_msssim_c240.csv};
        \addlegendentry{H.264 + LDPC 3/4 16QAM}
        
        \end{axis}
        \end{tikzpicture}
        }
  \caption{Performance comparison of DeepWiVe to H.264 paired with LDPC codes ($\rho=0.031$).}
  \label{fig:optimal_h264_kn0031} 
\end{figure}

\begin{figure} 
    \centering
  \subfloat[PSNR\label{subfig:psnr_optimal_h265_kn0031}]{%
    \begin{tikzpicture}
        \pgfplotsset{
            legend style={
                font=\fontsize{4}{4}\selectfont,
                at={(1.0,.0)},
                anchor=south east,
            },
            width=0.5\textwidth,
            xmin=-6,
            xmax=20,
            ymin=26,
            ymax=42,
            xtick distance=2,
            ytick distance=2,
            xlabel={SNR (dB)},
            ylabel={PSNR (dB)},
            grid=both,
            grid style={line width=.1pt, draw=gray!10},
            major grid style={line width=.2pt,draw=gray!50},
            every axis/.append style={
                x label style={
                    font=\fontsize{8}{8}\selectfont,
                    at={(axis description cs:0.5,-0.05)},
                    },
                y label style={
                    font=\fontsize{8}{8}\selectfont,
                    at={(axis description cs:-0.1,0.5)},
                    },
                x tick label style={
                    font=\fontsize{8}{8}\selectfont,
                    /pgf/number format/.cd,
                    fixed,
                    fixed zerofill,
                    precision=0,
                    /tikz/.cd
                    },
                y tick label style={
                    font=\fontsize{8}{8}\selectfont,
                    /pgf/number format/.cd,
                    fixed,
                    fixed zerofill,
                    precision=1,
                    /tikz/.cd
                    },
            }
        }
        \begin{axis}
        \addplot[cyan, solid, line width=0.9pt, mark=triangle*, mark options={fill=cyan, scale=1.6}, error bars/.cd, y dir=both, y explicit, every nth mark=2] 
                table [x=snrtest, y=PSNR, y error=PSNR_error, col sep=comma] {Data/deepwive_optimal_c240.csv};
        \addlegendentry{\textit{DeepWiVe} ($\text{SNR}_{\text{Est}}=\text{SNR}$)}
        
        \addplot[color=blue, dashed, line width=1.2pt, mark=*, mark options={fill=blue, solid, scale=1.1}, error bars/.cd, y dir=both, y explicit, every nth mark=2] 
        table [x=snrtest, y=bpsk, y error=bpsk_error, col sep=comma] {Data/h265_1,2LDPC_psnr_c240.csv};
        \addlegendentry{H.265 + LDPC 1/2 BPSK}
        
        \addplot[color=blue, dashed, line width=1.2pt, mark=triangle*, mark options={fill=blue, solid, scale=1.1}, error bars/.cd, y dir=both, y explicit, every nth mark=2] 
        table [x=snrtest, y=qpsk, y error=qpsk_error, col sep=comma] {Data/h265_1,2LDPC_psnr_c240.csv};
        \addlegendentry{H.265 + LDPC 1/2 QPSK}
        
        \addplot[color=blue, dashed, line width=1.2pt, mark=square*, mark options={fill=blue, solid, scale=1.1}, error bars/.cd, y dir=both, y explicit, every nth mark=2] 
        table [x=snrtest, y=16qam, y error=16qam_error, col sep=comma] {Data/h265_1,2LDPC_psnr_c240.csv};
        \addlegendentry{H.265 + LDPC 1/2 16QAM}
        
        
        \addplot[color=magenta, dashed, line width=1.2pt, mark=*, mark options={fill=magenta, solid, scale=1.1}, error bars/.cd, y dir=both, y explicit, every nth mark=2] 
        table [x=snrtest, y=bpsk, y error=bpsk_error, col sep=comma] {Data/h265_2,3LDPC_psnr_c240.csv};
        \addlegendentry{H.265 + LDPC 3/4 BPSK}
        
        \addplot[color=magenta, dashed, line width=1.2pt, mark=triangle*, mark options={fill=magenta, solid, scale=1.1}, error bars/.cd, y dir=both, y explicit, every nth mark=2] 
        table [x=snrtest, y=qpsk, y error=qpsk_error, col sep=comma] {Data/h265_2,3LDPC_psnr_c240.csv};
        \addlegendentry{H.265 + LDPC 3/4 QPSK}
        
        \addplot[color=magenta, dashed, line width=1.2pt, mark=square*, mark options={fill=magenta, solid, scale=1.1}, error bars/.cd, y dir=both, y explicit, every nth mark=2] 
        table [x=snrtest, y=16qam, y error=16qam_error, col sep=comma] {Data/h265_2,3LDPC_psnr_c240.csv};
        \addlegendentry{H.265 + LDPC 3/4 16QAM}
        
        \end{axis}
        \end{tikzpicture}
    }
    \\
  \subfloat[MS-SSIM\label{subfig:msssim_optimal_h265_kn0031}]{%
    \begin{tikzpicture}
        \pgfplotsset{
            legend style={
                font=\fontsize{4}{4}\selectfont,
                at={(1.0,0.)},
                anchor=south east,
            },
            width=0.5\textwidth,
            xmin=-6,
            xmax=20,
            ymin=0.94,
            ymax=1,
            xtick distance=2,
            ytick distance=0.01,
            xlabel={SNR (dB)},
            ylabel={MS-SSIM},
            grid=both,
            grid style={line width=.1pt, draw=gray!10},
            major grid style={line width=.2pt,draw=gray!50},
            every axis/.append style={
                x label style={
                    font=\fontsize{8}{8}\selectfont,
                    at={(axis description cs:0.5,-0.05)},
                    },
                y label style={
                    font=\fontsize{8}{8}\selectfont,
                    at={(axis description cs:-0.1,0.5)},
                    },
                x tick label style={
                    font=\fontsize{8}{8}\selectfont,
                    /pgf/number format/.cd,
                    fixed,
                    fixed zerofill,
                    precision=0,
                    /tikz/.cd
                    },
                y tick label style={
                    font=\fontsize{8}{8}\selectfont,
                    /pgf/number format/.cd,
                    fixed,
                    fixed zerofill,
                    precision=2,
                    /tikz/.cd
                    },
            }
        }
        \begin{axis}[mark options={solid}]
        \addplot[cyan, solid, line width=0.9pt, mark=triangle*, mark options={fill=cyan, scale=1.6}, error bars/.cd, y dir=both, y explicit, every nth mark=2] 
                table [x=snrtest, y=MS-SSIM, y error=MS-SSIM_error, col sep=comma] {Data/deepwive_optimal_c240.csv};
        \addlegendentry{\textit{DeepWiVe} ($\text{SNR}_{\text{Est}}=\text{SNR}$)}
        
        \addplot[color=blue, dashed, line width=1.2pt, mark=*, mark options={fill=blue, solid, scale=1.1}, error bars/.cd, y dir=both, y explicit, every nth mark=2] 
        table [x=snrtest, y=bpsk, y error=bpsk_error, col sep=comma] {Data/h265_1,2LDPC_msssim_c240.csv};
        \addlegendentry{H.265 + LDPC 1/2 BPSK}
        
        \addplot[color=blue, dashed, line width=1.2pt, mark=triangle*, mark options={fill=blue, solid, scale=1.1}, error bars/.cd, y dir=both, y explicit, every nth mark=2] 
        table [x=snrtest, y=qpsk, y error=qpsk_error, col sep=comma] {Data/h265_1,2LDPC_msssim_c240.csv};
        \addlegendentry{H.265 + LDPC 1/2 QPSK}
        
        \addplot[color=blue, dashed, line width=1.2pt, mark=square*, mark options={fill=blue, solid, scale=1.1}, error bars/.cd, y dir=both, y explicit, every nth mark=2] 
        table [x=snrtest, y=16qam, y error=16qam_error, col sep=comma] {Data/h265_1,2LDPC_msssim_c240.csv};
        \addlegendentry{H.265 + LDPC 1/2 16QAM}

        \addplot[color=magenta, dashed, line width=1.2pt, mark=*, mark options={fill=magenta, solid, scale=1.1}, error bars/.cd, y dir=both, y explicit, every nth mark=2] 
        table [x=snrtest, y=bpsk, y error=bpsk_error, col sep=comma] {Data/h265_2,3LDPC_msssim_c240.csv};
        \addlegendentry{H.265 + LDPC 3/4 BPSK}
        
        \addplot[color=magenta, dashed, line width=1.2pt, mark=triangle*, mark options={fill=magenta, solid, scale=1.1}, error bars/.cd, y dir=both, y explicit, every nth mark=2] 
        table [x=snrtest, y=qpsk, y error=qpsk_error, col sep=comma] {Data/h265_2,3LDPC_msssim_c240.csv};
        \addlegendentry{H.265 + LDPC 3/4 QPSK}
        
        \addplot[color=magenta, dashed, line width=1.2pt, mark=square*, mark options={fill=magenta, solid, scale=1.1}, error bars/.cd, y dir=both, y explicit, every nth mark=2] 
        table [x=snrtest, y=16qam, y error=16qam_error, col sep=comma] {Data/h265_2,3LDPC_msssim_c240.csv};
        \addlegendentry{H.265 + LDPC 3/4 16QAM}
        
        \end{axis}
        \end{tikzpicture}
        }
  \caption{Performance comparison of DeepWiVe to H.265 paired with LDPC codes ($\rho=0.031$).}
  \label{fig:optimal_h265_kn0031} 
\end{figure}
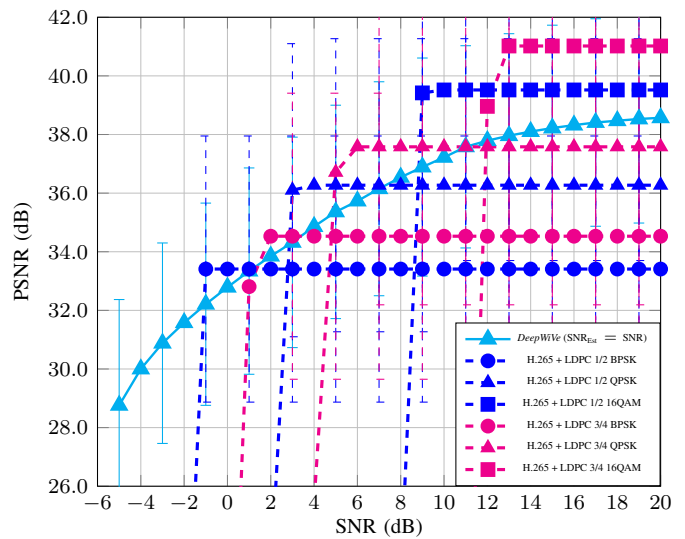
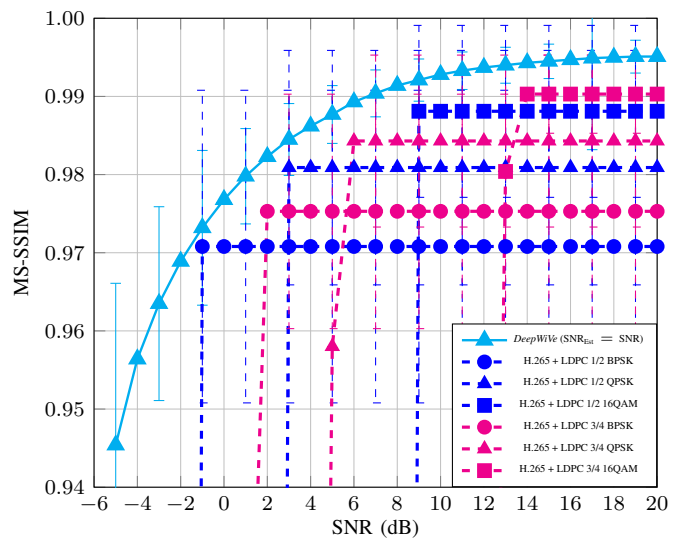

\begin{figure} 
    \centering
  \subfloat[PSNR\label{subfig:psnr_optimal_h264_kn0018}]{%
    \begin{tikzpicture}
        \pgfplotsset{
            legend style={
                font=\fontsize{4}{4}\selectfont,
                at={(1.0,.0)},
                anchor=south east,
            },
            width=0.5\textwidth,
            xmin=-6,
            xmax=20,
            ymin=16,
            ymax=37,
            xtick distance=2,
            ytick distance=2,
            xlabel={SNR (dB)},
            ylabel={PSNR (dB)},
            grid=both,
            grid style={line width=.1pt, draw=gray!10},
            major grid style={line width=.2pt,draw=gray!50},
            every axis/.append style={
                x label style={
                    font=\fontsize{8}{8}\selectfont,
                    at={(axis description cs:0.5,-0.05)},
                    },
                y label style={
                    font=\fontsize{8}{8}\selectfont,
                    at={(axis description cs:-0.1,0.5)},
                    },
                x tick label style={
                    font=\fontsize{8}{8}\selectfont,
                    /pgf/number format/.cd,
                    fixed,
                    fixed zerofill,
                    precision=0,
                    /tikz/.cd
                    },
                y tick label style={
                    font=\fontsize{8}{8}\selectfont,
                    /pgf/number format/.cd,
                    fixed,
                    fixed zerofill,
                    precision=1,
                    /tikz/.cd
                    },
            }
        }
        \begin{axis}
        \addplot[cyan, solid, line width=0.9pt, mark=triangle*, mark options={fill=cyan, scale=1.6}, error bars/.cd, y dir=both, y explicit, every nth mark=2] 
                table [x=snrtest, y=PSNR, y error=PSNR_error, col sep=comma] {Data/deepwive_optimal_c144.csv};
        \addlegendentry{\textit{DeepWiVe} ($\text{SNR}_{\text{Est}}=\text{SNR}$)}
        
        \addplot[color=blue, dashed, line width=1.2pt, mark=*, mark options={fill=blue, solid, scale=1.1}, error bars/.cd, y dir=both, y explicit, every nth mark=2] 
        table [x=snrtest, y=bpsk, y error=bpsk_error, col sep=comma] {Data/h264_1,2LDPC_psnr_c144.csv};
        \addlegendentry{H.264 + LDPC 1/2 BPSK}
        
        \addplot[color=blue, dashed, line width=1.2pt, mark=triangle*, mark options={fill=blue, solid, scale=1.1}, error bars/.cd, y dir=both, y explicit, every nth mark=2] 
        table [x=snrtest, y=qpsk, y error=qpsk_error, col sep=comma] {Data/h264_1,2LDPC_psnr_c144.csv};
        \addlegendentry{H.264 + LDPC 1/2 QPSK}
        
        \addplot[color=blue, dashed, line width=1.2pt, mark=square*, mark options={fill=blue, solid, scale=1.1}, error bars/.cd, y dir=both, y explicit, every nth mark=2] 
        table [x=snrtest, y=16qam, y error=16qam_error, col sep=comma] {Data/h264_1,2LDPC_psnr_c144.csv};
        \addlegendentry{H.264 + LDPC 1/2 16QAM}
        
        
        \addplot[color=magenta, dashed, line width=1.2pt, mark=*, mark options={fill=magenta, solid, scale=1.1}, error bars/.cd, y dir=both, y explicit, every nth mark=2] 
        table [x=snrtest, y=bpsk, y error=bpsk_error, col sep=comma] {Data/h264_2,3LDPC_psnr_c144.csv};
        \addlegendentry{H.264 + LDPC 3/4 BPSK}
        
        \addplot[color=magenta, dashed, line width=1.2pt, mark=triangle*, mark options={fill=magenta, solid, scale=1.1}, error bars/.cd, y dir=both, y explicit, every nth mark=2] 
        table [x=snrtest, y=qpsk, y error=qpsk_error, col sep=comma] {Data/h264_2,3LDPC_psnr_c144.csv};
        \addlegendentry{H.264 + LDPC 3/4 QPSK}
        
        \addplot[color=magenta, dashed, line width=1.2pt, mark=square*, mark options={fill=magenta, solid, scale=1.1}, error bars/.cd, y dir=both, y explicit, every nth mark=2] 
        table [x=snrtest, y=16qam, y error=16qam_error, col sep=comma] {Data/h264_2,3LDPC_psnr_c144.csv};
        \addlegendentry{H.264 + LDPC 3/4 16QAM}
        
        \end{axis}
        \end{tikzpicture}
    }
    \\
  \subfloat[MS-SSIM\label{subfig:msssim_optimal_h264_kn0018}]{%
    \begin{tikzpicture}
        \pgfplotsset{
            legend style={
                font=\fontsize{4}{4}\selectfont,
                at={(1.0,0.)},
                anchor=south east,
            },
            width=0.5\textwidth,
            xmin=-6,
            xmax=20,
            ymin=0.74,
            ymax=1,
            xtick distance=2,
            ytick distance=0.02,
            xlabel={SNR (dB)},
            ylabel={MS-SSIM},
            grid=both,
            grid style={line width=.1pt, draw=gray!10},
            major grid style={line width=.2pt,draw=gray!50},
            every axis/.append style={
                x label style={
                    font=\fontsize{8}{8}\selectfont,
                    at={(axis description cs:0.5,-0.05)},
                    },
                y label style={
                    font=\fontsize{8}{8}\selectfont,
                    at={(axis description cs:-0.1,0.5)},
                    },
                x tick label style={
                    font=\fontsize{8}{8}\selectfont,
                    /pgf/number format/.cd,
                    fixed,
                    fixed zerofill,
                    precision=0,
                    /tikz/.cd
                    },
                y tick label style={
                    font=\fontsize{8}{8}\selectfont,
                    /pgf/number format/.cd,
                    fixed,
                    fixed zerofill,
                    precision=2,
                    /tikz/.cd
                    },
            }
        }
        \begin{axis}[mark options={solid}]
        \addplot[cyan, solid, line width=0.9pt, mark=triangle*, mark options={fill=cyan, scale=1.6}, error bars/.cd, y dir=both, y explicit, every nth mark=2] 
                table [x=snrtest, y=MS-SSIM, y error=MS-SSIM_error, col sep=comma] {Data/deepwive_optimal_c144.csv};
        \addlegendentry{\textit{DeepWiVe} ($\text{SNR}_{\text{Est}}=\text{SNR}$)}
        
        \addplot[color=blue, dashed, line width=1.2pt, mark=*, mark options={fill=blue, solid, scale=1.1}, error bars/.cd, y dir=both, y explicit, every nth mark=2] 
        table [x=snrtest, y=bpsk, y error=bpsk_error, col sep=comma] {Data/h264_1,2LDPC_msssim_c144.csv};
        \addlegendentry{H.264 + LDPC 1/2 BPSK}
        
        \addplot[color=blue, dashed, line width=1.2pt, mark=triangle*, mark options={fill=blue, solid, scale=1.1}, error bars/.cd, y dir=both, y explicit, every nth mark=2] 
        table [x=snrtest, y=qpsk, y error=qpsk_error, col sep=comma] {Data/h264_1,2LDPC_msssim_c144.csv};
        \addlegendentry{H.264 + LDPC 1/2 QPSK}
        
        \addplot[color=blue, dashed, line width=1.2pt, mark=square*, mark options={fill=blue, solid, scale=1.1}, error bars/.cd, y dir=both, y explicit, every nth mark=2] 
        table [x=snrtest, y=16qam, y error=16qam_error, col sep=comma] {Data/h264_1,2LDPC_msssim_c144.csv};
        \addlegendentry{H.264 + LDPC 1/2 16QAM}
        
        
        \addplot[color=magenta, dashed, line width=1.2pt, mark=*, mark options={fill=magenta, solid, scale=1.1}, error bars/.cd, y dir=both, y explicit, every nth mark=2] 
        table [x=snrtest, y=bpsk, y error=bpsk_error, col sep=comma] {Data/h264_2,3LDPC_msssim_c144.csv};
        \addlegendentry{H.264 + LDPC 3/4 BPSK}
        
        \addplot[color=magenta, dashed, line width=1.2pt, mark=triangle*, mark options={fill=magenta, solid, scale=1.1}, error bars/.cd, y dir=both, y explicit, every nth mark=2] 
        table [x=snrtest, y=qpsk, y error=qpsk_error, col sep=comma] {Data/h264_2,3LDPC_msssim_c144.csv};
        \addlegendentry{H.264 + LDPC 3/4 QPSK}
        
        \addplot[color=magenta, dashed, line width=1.2pt, mark=square*, mark options={fill=magenta, solid, scale=1.1}, error bars/.cd, y dir=both, y explicit, every nth mark=2] 
        table [x=snrtest, y=16qam, y error=16qam_error, col sep=comma] {Data/h264_2,3LDPC_msssim_c144.csv};
        \addlegendentry{H.264 + LDPC 3/4 16QAM}
        
        \end{axis}
        \end{tikzpicture}
        }
  \caption{Performance comparison of DeepWiVe to H.264 paired with LDPC codes ($\rho=0.018$).}
  \label{fig:optimal_h264_kn0018} 
\end{figure}
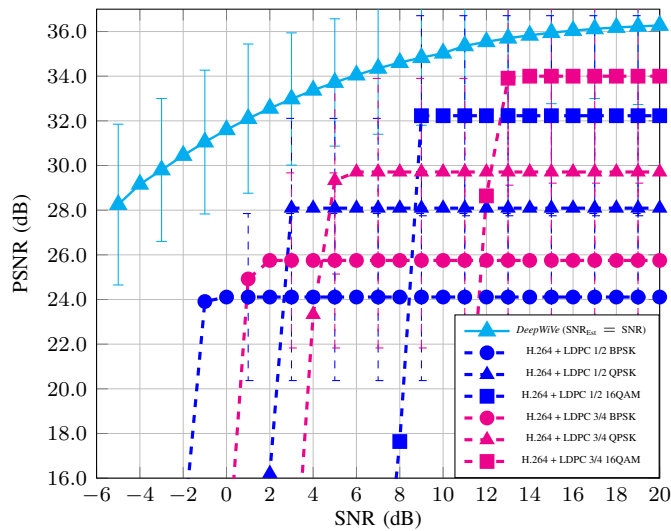
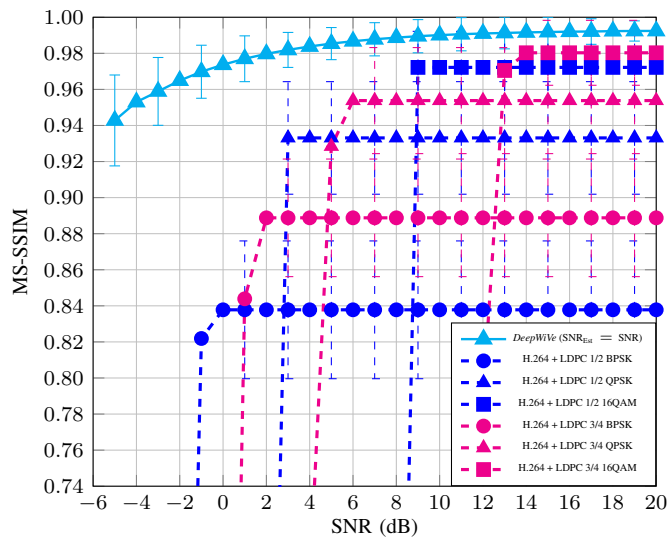

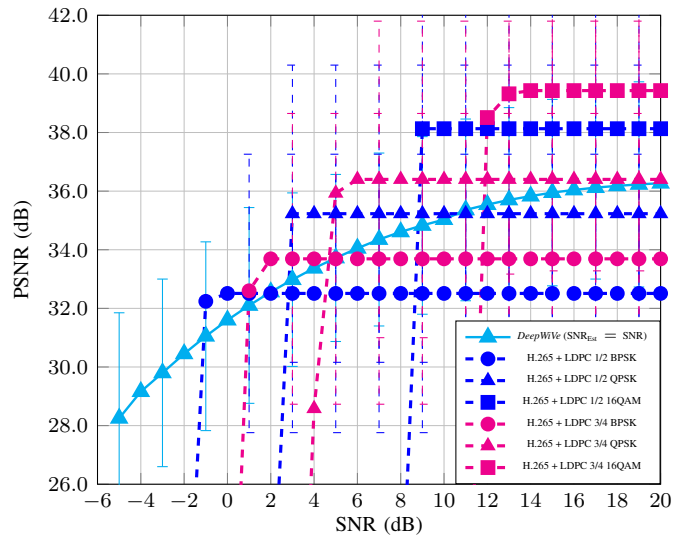
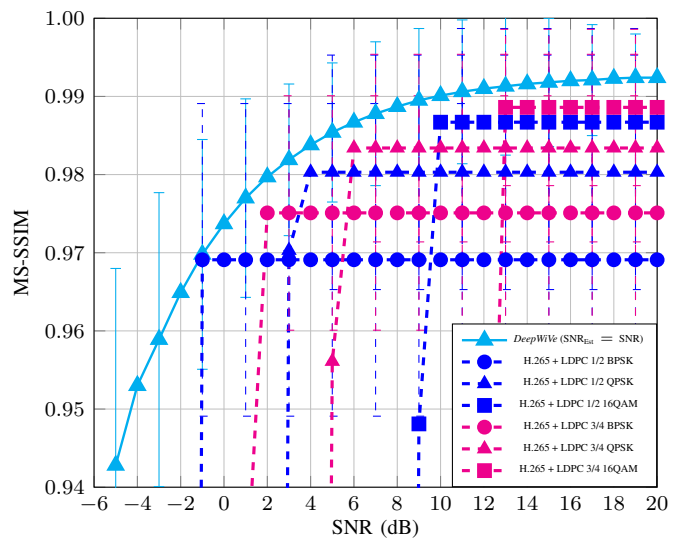
\begin{figure} 
    \centering
  \subfloat[PSNR\label{subfig:psnr_optimal_h265_kn0018}]{%
    \begin{tikzpicture}
        \pgfplotsset{
            legend style={
                font=\fontsize{4}{4}\selectfont,
                at={(1.0,.0)},
                anchor=south east,
            },
            width=0.5\textwidth,
            xmin=-6,
            xmax=20,
            ymin=26,
            ymax=42,
            xtick distance=2,
            ytick distance=2,
            xlabel={SNR (dB)},
            ylabel={PSNR (dB)},
            grid=both,
            grid style={line width=.1pt, draw=gray!10},
            major grid style={line width=.2pt,draw=gray!50},
            every axis/.append style={
                x label style={
                    font=\fontsize{8}{8}\selectfont,
                    at={(axis description cs:0.5,-0.05)},
                    },
                y label style={
                    font=\fontsize{8}{8}\selectfont,
                    at={(axis description cs:-0.1,0.5)},
                    },
                x tick label style={
                    font=\fontsize{8}{8}\selectfont,
                    /pgf/number format/.cd,
                    fixed,
                    fixed zerofill,
                    precision=0,
                    /tikz/.cd
                    },
                y tick label style={
                    font=\fontsize{8}{8}\selectfont,
                    /pgf/number format/.cd,
                    fixed,
                    fixed zerofill,
                    precision=1,
                    /tikz/.cd
                    },
            }
        }
        \begin{axis}
        \addplot[cyan, solid, line width=0.9pt, mark=triangle*, mark options={fill=cyan, scale=1.6}, error bars/.cd, y dir=both, y explicit, every nth mark=2] 
                table [x=snrtest, y=PSNR, y error=PSNR_error, col sep=comma] {Data/deepwive_optimal_c144.csv};
        \addlegendentry{\textit{DeepWiVe} ($\text{SNR}_{\text{Est}}=\text{SNR}$)}
        
        \addplot[color=blue, dashed, line width=1.2pt, mark=*, mark options={fill=blue, solid, scale=1.1}, error bars/.cd, y dir=both, y explicit, every nth mark=2] 
        table [x=snrtest, y=bpsk, y error=bpsk_error, col sep=comma] {Data/h265_1,2LDPC_psnr_c144.csv};
        \addlegendentry{H.265 + LDPC 1/2 BPSK}
        
        \addplot[color=blue, dashed, line width=1.2pt, mark=triangle*, mark options={fill=blue, solid, scale=1.1}, error bars/.cd, y dir=both, y explicit, every nth mark=2] 
        table [x=snrtest, y=qpsk, y error=qpsk_error, col sep=comma] {Data/h265_1,2LDPC_psnr_c144.csv};
        \addlegendentry{H.265 + LDPC 1/2 QPSK}
        
        \addplot[color=blue, dashed, line width=1.2pt, mark=square*, mark options={fill=blue, solid, scale=1.1}, error bars/.cd, y dir=both, y explicit, every nth mark=2] 
        table [x=snrtest, y=16qam, y error=16qam_error, col sep=comma] {Data/h265_1,2LDPC_psnr_c144.csv};
        \addlegendentry{H.265 + LDPC 1/2 16QAM}
        
        
        \addplot[color=magenta, dashed, line width=1.2pt, mark=*, mark options={fill=magenta, solid, scale=1.1}, error bars/.cd, y dir=both, y explicit, every nth mark=2] 
        table [x=snrtest, y=bpsk, y error=bpsk_error, col sep=comma] {Data/h265_2,3LDPC_psnr_c144.csv};
        \addlegendentry{H.265 + LDPC 3/4 BPSK}
        
        \addplot[color=magenta, dashed, line width=1.2pt, mark=triangle*, mark options={fill=magenta, solid, scale=1.1}, error bars/.cd, y dir=both, y explicit, every nth mark=2] 
        table [x=snrtest, y=qpsk, y error=qpsk_error, col sep=comma] {Data/h265_2,3LDPC_psnr_c144.csv};
        \addlegendentry{H.265 + LDPC 3/4 QPSK}
        
        \addplot[color=magenta, dashed, line width=1.2pt, mark=square*, mark options={fill=magenta, solid, scale=1.1}, error bars/.cd, y dir=both, y explicit, every nth mark=2] 
        table [x=snrtest, y=16qam, y error=16qam_error, col sep=comma] {Data/h265_2,3LDPC_psnr_c144.csv};
        \addlegendentry{H.265 + LDPC 3/4 16QAM}
        
        \end{axis}
        \end{tikzpicture}
    }
    \\
  \subfloat[MS-SSIM\label{subfig:msssim_optimal_h265_kn0018}]{%
    \begin{tikzpicture}
        \pgfplotsset{
            legend style={
                font=\fontsize{4}{4}\selectfont,
                at={(1.0,0.)},
                anchor=south east,
            },
            width=0.5\textwidth,
            xmin=-6,
            xmax=20,
            ymin=0.94,
            ymax=1,
            xtick distance=2,
            ytick distance=0.01,
            xlabel={SNR (dB)},
            ylabel={MS-SSIM},
            grid=both,
            grid style={line width=.1pt, draw=gray!10},
            major grid style={line width=.2pt,draw=gray!50},
            every axis/.append style={
                x label style={
                    font=\fontsize{8}{8}\selectfont,
                    at={(axis description cs:0.5,-0.05)},
                    },
                y label style={
                    font=\fontsize{8}{8}\selectfont,
                    at={(axis description cs:-0.1,0.5)},
                    },
                x tick label style={
                    font=\fontsize{8}{8}\selectfont,
                    /pgf/number format/.cd,
                    fixed,
                    fixed zerofill,
                    precision=0,
                    /tikz/.cd
                    },
                y tick label style={
                    font=\fontsize{8}{8}\selectfont,
                    /pgf/number format/.cd,
                    fixed,
                    fixed zerofill,
                    precision=2,
                    /tikz/.cd
                    },
            }
        }
        \begin{axis}[mark options={solid}]
        \addplot[cyan, solid, line width=0.9pt, mark=triangle*, mark options={fill=cyan, scale=1.6}, error bars/.cd, y dir=both, y explicit, every nth mark=2] 
                table [x=snrtest, y=MS-SSIM, y error=MS-SSIM_error, col sep=comma] {Data/deepwive_optimal_c144.csv};
        \addlegendentry{\textit{DeepWiVe} ($\text{SNR}_{\text{Est}}=\text{SNR}$)}
        
        \addplot[color=blue, dashed, line width=1.2pt, mark=*, mark options={fill=blue, solid, scale=1.1}, error bars/.cd, y dir=both, y explicit, every nth mark=2] 
        table [x=snrtest, y=bpsk, y error=bpsk_error, col sep=comma] {Data/h265_1,2LDPC_msssim_c144.csv};
        \addlegendentry{H.265 + LDPC 1/2 BPSK}
        
        \addplot[color=blue, dashed, line width=1.2pt, mark=triangle*, mark options={fill=blue, solid, scale=1.1}, error bars/.cd, y dir=both, y explicit, every nth mark=2] 
        table [x=snrtest, y=qpsk, y error=qpsk_error, col sep=comma] {Data/h265_1,2LDPC_msssim_c144.csv};
        \addlegendentry{H.265 + LDPC 1/2 QPSK}
        
        \addplot[color=blue, dashed, line width=1.2pt, mark=square*, mark options={fill=blue, solid, scale=1.1}, error bars/.cd, y dir=both, y explicit, every nth mark=2] 
        table [x=snrtest, y=16qam, y error=16qam_error, col sep=comma] {Data/h265_1,2LDPC_msssim_c144.csv};
        \addlegendentry{H.265 + LDPC 1/2 16QAM}
        
        
        \addplot[color=magenta, dashed, line width=1.2pt, mark=*, mark options={fill=magenta, solid, scale=1.1}, error bars/.cd, y dir=both, y explicit, every nth mark=2] 
        table [x=snrtest, y=bpsk, y error=bpsk_error, col sep=comma] {Data/h265_2,3LDPC_msssim_c144.csv};
        \addlegendentry{H.265 + LDPC 3/4 BPSK}
        
        \addplot[color=magenta, dashed, line width=1.2pt, mark=triangle*, mark options={fill=magenta, solid, scale=1.1}, error bars/.cd, y dir=both, y explicit, every nth mark=2] 
        table [x=snrtest, y=qpsk, y error=qpsk_error, col sep=comma] {Data/h265_2,3LDPC_msssim_c144.csv};
        \addlegendentry{H.265 + LDPC 3/4 QPSK}
        
        \addplot[color=magenta, dashed, line width=1.2pt, mark=square*, mark options={fill=magenta, solid, scale=1.1}, error bars/.cd, y dir=both, y explicit, every nth mark=2] 
        table [x=snrtest, y=16qam, y error=16qam_error, col sep=comma] {Data/h265_2,3LDPC_msssim_c144.csv};
        \addlegendentry{H.265 + LDPC 3/4 16QAM}
        
        \end{axis}
        \end{tikzpicture}
        }
  \caption{Performance comparison of DeepWiVe to H.265 paired with LDPC codes ($\rho=0.018$).}
  \label{fig:optimal_h265_kn0018} 
\end{figure}

\subsection{Simulation Results}
\label{subsec:num_results}

We compare the performance of our model with that of the conventional separation-based schemes.
In particular, we use the H.264 \cite{wiegand_overview_2003} and H.265 \cite{ohm_high_2013} video compression codecs for source coding, LDPC codes \cite{gallager_low-density_1962} for channel coding, and QAM modulation.
We plot the average video quality across the test dataset using each of the schemes considered herein and error bars representing the standard deviation of the video qualities.

In Fig. \ref{fig:graceful_h264_kn0031}, we show the effect of channel estimation error on the performance of \textit{DeepWiVe}.
We specifically compare models using $\text{SNR}_{\text{Est}}$ where the H.264 codec paired with a specific LDPC code rate experiences the \textit{cliff-effect}.
It is clear that \textit{DeepWiVe} is able to overcome the \textit{cliff-effect}, as video quality gracefully degrades as the SNR decreases even as $\text{SNR}_{\text{Est}}$ remains the same.
This is in contrast to the cliff edge drop off that separation-based designs suffer from.
The \textit{cliff-effect} is due to the fact that as the SNR decreases, the capacity of the channel decreases, leading to certain LDPC rate and modulation order pairs to communicate at a rate greater than the capacity of the channel, which leads to highly unreliable communication, manifested as a cliff edge deterioration of the video quality. 
We can also see that the variations in the video quality using \textit{DeepWiVe} is lower than those produced by the separation scheme as indicated by the smaller error bars.
The error bars represent the standard deviation of the video quality at the receiver side.
This is likely due to the fact that the H.264 codec does not have a continuous range of compression rates available but rather a set of discrete levels it can compress.
Depending on the complexity of the video, a given target distortion may lead to a larger rate than is allowed by the instantaneous channel condition and it must reduce the target distortion level.
Since the allowed target distortion levels can be far apart, this may mean the video is compressed more conservatively than is suggested by the channel in order to meet the channel condition, leading to a large variation in the resultant video quality.
\textit{DeepWiVe}, on the other hand, does not have this issue as we do not define a set of possible compression rates.
Instead, the weights are adjusted by the AF modules based on the current channel condition to meet the rate-distortion curve as closely as possible.
Fig. \ref{fig:example_frames} shows the visual results of the plots shown in Fig. \ref{fig:graceful_h264_kn0031} for a specific video in the test dataset.
It can be seen that at $\text{SNR}=13$dB, the visual qualities of the videos produced by H.264 and \textit{DeepWiVe} are similar.
However, at $\text{SNR}=3$dB, the video produced by H.264 starts to look very pixelated, while \textit{DeepWiVe} is still able to retain a smooth looking frame.
At $\text{SNR}=-4$dB, the capacity of the channel is too low for H.264 to compress the video sufficiently, therefore the output is simply black, while \textit{DeepWiVe} is still able to achieve a reasonable video quality despite the very low channel SNR.

In Fig. \ref{fig:optimal_h264_kn0031}, we present the comparison of \textit{DeepWiVe} using the accurate estimate of the channel SNR (i.e., $\text{SNR}_{\text{Est}}=\text{SNR}$) with separation employing H.264, and in Fig. \ref{fig:optimal_h265_kn0031} employing H.265.
In Fig. \ref{fig:optimal_h264_kn0031}, we see that at $\rho=0.031$, \textit{DeepWiVe} is superior to the separation based scheme using H.264 in all the SNRs tested.
This is due to the end-to-end optimization of the JSCC encoder and decoder producing a superior compression, channel code, and modulation scheme than the separation-based scheme.
When compared to H.265, we see in Fig. \ref{subfig:psnr_optimal_h265_kn0031} that H.265 outperforms \textit{DeepWiVe} in terms of the PSNR metric.
However, when compared with the more perceptually aligned MS-SSIM metric in Fig. \ref{subfig:msssim_optimal_h265_kn0031}, we see that \textit{DeepWiVe} can also outperform separation-based transmission with H.265.
We also highlight that, in the very low SNR regime (i.e., SNR $<-1$dB), H.265 was unable to meet the compression rate required, and therefore did not produce results in that rage.
\textit{DeepWiVe} on the other hand, did not have this problem.
We believe that further optimization of the network architecture can bring \textit{DeepWiVe} on par or surpass H.265 evaluated using the PSNR metric for higher SNR values as well.
On average, \textit{DeepWiVe} is 1.51dB better in PSNR and 0.0088 better in MS-SSIM than H.264 for $\text{SNR}\in[13,20]$dB, 3.61dB better in PSNR and 0.0281 better in MS-SSIM for $\text{SNR}\in[3,6]$dB.
For H.265, \textit{DeepWiVe} is 2.69dB worse in PSNR and 0.0056 better in MS-SSIM for $\text{SNR}\in[13,20]$dB, 1.59dB worse in PSNR and 0.0109 better in MS-SSIM for $\text{SNR}\in[3,6]$dB.

Next, we investigate the variable bandwidth transmission capability of \textit{DeepWiVe} by decreasing the bandwidth compression ratio $\rho$, thereby increasing the compression of the video. 
To decrease $\rho$, we do not require the retraining of the autoencoder networks ($f_{\boldsymbol{\theta}}, f_{\boldsymbol{\theta}^\prime}, g_{\boldsymbol{\phi}}, g_{\boldsymbol{\phi}^\prime}, h_{\boldsymbol{\eta}}$); 
we only need to retrain the bandwidth allocator $q_{\boldsymbol{\psi}}$.
As shown in Fig. \ref{fig:optimal_h264_kn0018}, we see that \textit{DeepWiVe} beats H.264 with LDPC coding for all SNRs tested in terms of both the PSNR and MS-SSIM metrics. 
It also beats H.265 using the MS-SSIM metric as shown in Fig. \ref{subfig:msssim_optimal_h265_kn0018}, although again, it falls short of H.265 in terms of the PSNR metric for SNR $>-1$ dB (Fig. \ref{subfig:psnr_optimal_h265_kn0018}).
This shows that \textit{DeepWiVe} can achieve variable bandwidth transmission using RL to allocate an arbitrary number of blocks to meet the desired transmission bandwidth, as outlined in Section \ref{subsec:bw_alloc}.
On average, for $\rho=0.018$, \textit{DeepWiVe} is 2.05 dB better in PSNR and 0.0129 better in MS-SSIM than H.264 for $\text{SNR}\in[13,20]$dB, 4.72dB better in PSNR and 0.0462 better in MS-SSIM for $\text{SNR}\in[3,6]$dB.
For H.265, \textit{DeepWiVe} is 3.37dB worse in PSNR and 0.0053 better in MS-SSIM for $\text{SNR}\in[13,20]$dB, 3.16dB worse in PSNR and 0.0058 better in MS-SSIM for $\text{SNR}\in[3,6]$dB.

\begin{figure} 
    \centering
  \subfloat[PSNR \label{subfig:uniform_v_optimal_psnr}]{%
    \begin{tikzpicture}
        \pgfplotsset{
            legend style={
                font=\fontsize{4}{4}\selectfont,
                at={(1.0,.0)},
                anchor=south east,
            },
            width=0.5\textwidth,
            xmin=-6,
            xmax=20,
            ymin=28,
            ymax=40,
            xtick distance=2,
            ytick distance=2,
            xlabel={SNR (dB)},
            ylabel={PSNR (dB)},
            grid=both,
            grid style={line width=.1pt, draw=gray!10},
            major grid style={line width=.2pt,draw=gray!50},
            every axis/.append style={
                x label style={
                    font=\fontsize{8}{8}\selectfont,
                    at={(axis description cs:0.5,-0.05)},
                    },
                y label style={
                    font=\fontsize{8}{8}\selectfont,
                    at={(axis description cs:-0.1,0.5)},
                    },
                x tick label style={
                    font=\fontsize{8}{8}\selectfont,
                    /pgf/number format/.cd,
                    fixed,
                    fixed zerofill,
                    precision=0,
                    /tikz/.cd
                    },
                y tick label style={
                    font=\fontsize{8}{8}\selectfont,
                    /pgf/number format/.cd,
                    fixed,
                    fixed zerofill,
                    precision=1,
                    /tikz/.cd
                    },
            }
        }
        \begin{axis}
        \addplot[cyan, solid, line width=0.9pt, mark=triangle*, mark options={fill=cyan, scale=1.6}, error bars/.cd, y dir=both, y explicit, every nth mark=2] 
                table [x=snrtest, y=PSNR, y error=PSNR_error, col sep=comma] {Data/deepwive_uniform_c240.csv};
        \addlegendentry{\textit{DeepWiVe} uniform ($\rho=0.031$)}
        
        \addplot[magenta, dashed, line width=1.2pt, mark=triangle*, mark options={fill=magenta, solid, scale=1.6}, error bars/.cd, y dir=both, y explicit, every nth mark=2] 
                table [x=snrtest, y=PSNR, y error=PSNR_error, col sep=comma] {Data/deepwive_optimal_c240.csv};
        \addlegendentry{\textit{DeepWiVe} Optimized ($\rho=0.031$)}
        
        \addplot[cyan, solid, line width=0.9pt, mark=square*, mark options={fill=cyan, scale=1.2}, error bars/.cd, y dir=both, y explicit, every nth mark=2] 
                table [x=snrtest, y=PSNR, y error=PSNR_error, col sep=comma] {Data/deepwive_uniform_c144.csv};
        \addlegendentry{\textit{DeepWiVe} uniform ($\rho=0.018$)}
        
        \addplot[magenta, dashed, line width=1.2pt, mark=square*, mark options={fill=magenta, solid, scale=1.2}, error bars/.cd, y dir=both, y explicit, every nth mark=2] 
                table [x=snrtest, y=PSNR, y error=PSNR_error, col sep=comma] {Data/deepwive_optimal_c144.csv};
        \addlegendentry{\textit{DeepWiVe} Optimized ($\rho=0.018$)}

        \end{axis}
        \end{tikzpicture}
    }
    \\
  \subfloat[MS-SSIM \label{subfig:uniform_v_optimal_msssim}]{%
    \begin{tikzpicture}
        \pgfplotsset{
            legend style={
                font=\fontsize{4}{4}\selectfont,
                at={(1.0,0.)},
                anchor=south east,
            },
            width=0.5\textwidth,
            xmin=-6,
            xmax=20,
            ymin=0.94,
            ymax=0.997,
            xtick distance=2,
            ytick distance=0.01,
            xlabel={SNR (dB)},
            ylabel={MS-SSIM},
            grid=both,
            grid style={line width=.1pt, draw=gray!10},
            major grid style={line width=.2pt,draw=gray!50},
            every axis/.append style={
                x label style={
                    font=\fontsize{8}{8}\selectfont,
                    at={(axis description cs:0.5,-0.05)},
                    },
                y label style={
                    font=\fontsize{8}{8}\selectfont,
                    at={(axis description cs:-0.1,0.5)},
                    },
                x tick label style={
                    font=\fontsize{8}{8}\selectfont,
                    /pgf/number format/.cd,
                    fixed,
                    fixed zerofill,
                    precision=0,
                    /tikz/.cd
                    },
                y tick label style={
                    font=\fontsize{8}{8}\selectfont,
                    /pgf/number format/.cd,
                    fixed,
                    fixed zerofill,
                    precision=2,
                    /tikz/.cd
                    },
            }
        }
        \begin{axis}[mark options={solid}]
        \addplot[cyan, solid, line width=0.9pt, mark=triangle*, mark options={fill=cyan, scale=1.6}, error bars/.cd, y dir=both, y explicit, every nth mark=2] 
                table [x=snrtest, y=MS-SSIM, y error=MS-SSIM_error, col sep=comma] {Data/deepwive_uniform_c240.csv};
        \addlegendentry{\textit{DeepWiVe} uniform ($\rho=0.031$)}
        
        \addplot[magenta, dashed, line width=1.2pt, mark=triangle*, mark options={fill=magenta, solid, scale=1.6}, error bars/.cd, y dir=both, y explicit, every nth mark=2] 
                table [x=snrtest, y=MS-SSIM, y error=MS-SSIM_error, col sep=comma] {Data/deepwive_optimal_c240.csv};
        \addlegendentry{\textit{DeepWiVe} Optimized ($\rho=0.031$)}
        
        \addplot[cyan, solid, line width=0.9pt, mark=square*, mark options={fill=cyan, scale=1.2}, error bars/.cd, y dir=both, y explicit, every nth mark=2] 
                table [x=snrtest, y=MS-SSIM, y error=MS-SSIM_error, col sep=comma] {Data/deepwive_uniform_c144.csv};
        \addlegendentry{\textit{DeepWiVe} uniform ($\rho=0.018$)}
        
        \addplot[magenta, dashed, line width=1.2pt, mark=square*, mark options={fill=magenta, solid, scale=1.2}, error bars/.cd, y dir=both, y explicit, every nth mark=2] 
                table [x=snrtest, y=MS-SSIM, y error=MS-SSIM_error, col sep=comma] {Data/deepwive_optimal_c144.csv};
        \addlegendentry{\textit{DeepWiVe} Optimized ($\rho=0.018$)}
        
        \end{axis}
        \end{tikzpicture}
        }
    \caption{Comparison of uniform bandwidth allocation versus optimal bandwidth allocation via auxiliary bandwidth allocation network $q_{\boldsymbol{\psi}}$.}
  \label{fig:uniform_v_optimal} 
\end{figure}
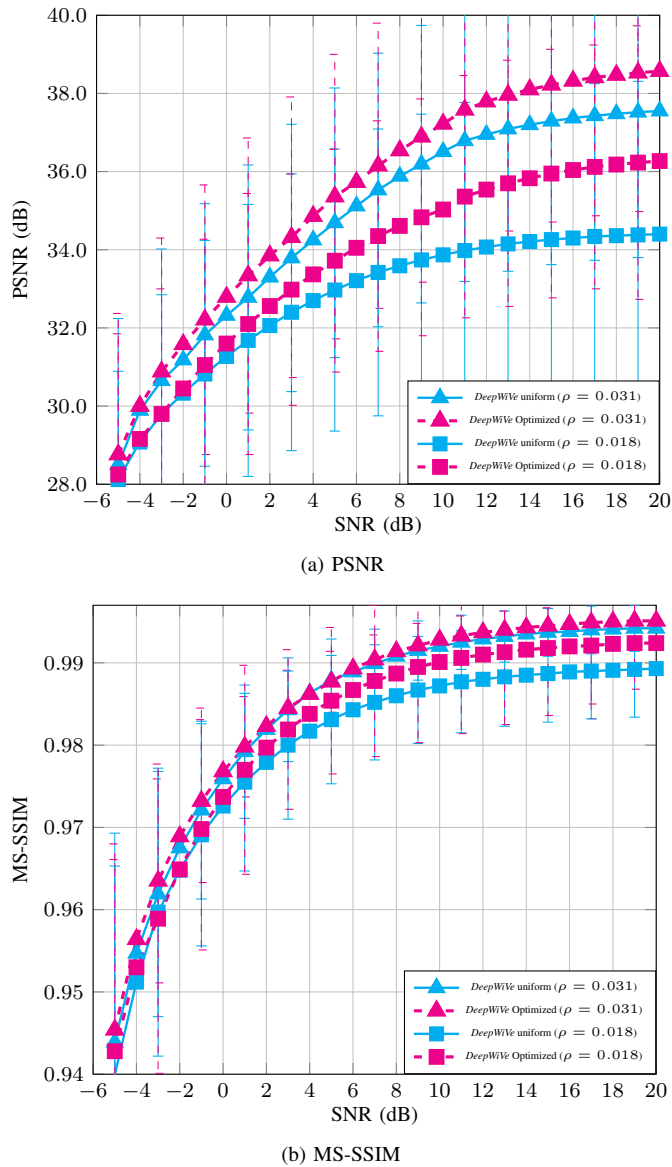

Lastly, we evaluate the performance of our models with and without the optimal allocation of bandwidth, as an ablation study. 
We compare the results obtained by using the allocation network $q_{\boldsymbol{\psi}}$ with that of uniform allocation (i.e., $v_i^n=5,~\forall i,n$ for $\rho=0.031$ and $v_i^n=3,~\forall i,n$ for $\rho=0.018$).
In Fig. \ref{fig:uniform_v_optimal}, it can be seen that there is a clear and significant improvement in performance when the allocation is optimized by our allocation network $q_{\boldsymbol{\psi}}$, by $0.67$ dB and $0.0008$ in PSNR and MS-SSIM, respectively, for $\rho=0.031$. 
For $\rho=0.018$, the gains are $0.99$dB and $0.0023$ in PSNR and MS-SSIM, respectively.
We observe that the gain from bandwidth allocation is more significant when $\rho$ is smaller; that is, when the available channel bandwidth is more limited.

\section{Conclusion}
\label{Sec:conclusion}

We presented the first ever DNN-aided joint source-channel wireless video transmission scheme in the literature. Our novel architecture, called \textit{DeepWiVe}, is capable of dynamic bandwidth allocation and residual estimation without the need for distortion feedback. Additionally, it utilizes RL to learn a bandwidth allocation network that optimizes the allocation of available bandwidth within a given GoP in a dynamic fashion with the goal of maximizing the visual quality of the video under the given bandwidth constraint. 
Our results show that \textit{DeepWiVe} overcomes the \textit{cliff-effect} that all separation-based schemes suffer from, and achieves a graceful degradation with channel quality. In highly bandwidth constrained scenarios, \textit{DeepWiVe}  produces far superior video quality compared to both H.264 and H.265.
We also show that our bandwidth allocation strategy is effective, improving upon the na\"ive uniform allocation by up to $0.87$dB in PSNR. 
Our overall results also show that \textit{DeepWiVe} is better than the separation-based schemes using industry standard H.264 codec and LDPC channel codes in all the channel conditions considered by up to $4.72$dB.
Although, H.265 performs better than \textit{DeepWiVe} in terms of the PSNR metric, \textit{DeepWiVe} outperforms H.265 when compared in terms of MS-SSIM, which is widely accepted as a performance measure that better represents human perceptual quality.

\bibliographystyle{ieeetr}
\bibliography{DeepJSCC-v.bib}

\end{document}